\documentclass[11pt,preprint]{aastex}

\shortauthors{Sluis \& Williams}
\shorttitle{Uncovering Planetary Nebulae}

\newcommand{\bm}[1]{\ensuremath{ \mathbf{#1} }}

\newcommand{\mr}[1]{\ensuremath{ \mathrm{#1} }}
\newcommand{\dl}[0]{\ensuremath{ \mathrm{d} }}

\begin{document}

\title{Uncovering Planetary Nebulae in Early Type Galaxies using
the Rutgers Fabry-Perot} 

\author{Arend P. N. Sluis\altaffilmark{1,2} 
        and T.B. Williams\altaffilmark{1}}

\affil{Department of Physics \& Astronomy, Rutgers University, \\ 
       136 Frelinghuysen Road, Piscataway, NJ 08854}

\email{sluis@physics.rutgers.edu, williams@physics.rutgers.edu}

\altaffiltext{1}{Visiting Astronomer, Cerro Telolo Inter-American
Observatory. CTIO is operated by AURA, Inc.\ under contract to the
National Science Foundation.}

\altaffiltext{2}{Current address: Dept. of Astronomy, University of
Massachusetts, LGRT-B 619E, 710 N.\ Pleasant St., Amherst, MA 01003;
email: sluis@astro.umass.edu.}

\begin{abstract}
We report on observations of four early-type galaxies performed with
the Rutgers Fabry-Perot in order to search for Planetary Nebulae (PNe)
in these systems.  The aim is to use the PNe as kinematic tracers of
the galaxy potential.  We describe our data reduction and analysis
procedure and show that the proper calibration of our detection
statistic is crucial in getting down to our limiting magnitude of
$m_{5007} = 26.1$.  In the case of the two Leo galaxies we find
moderately sized samples: 54 PNe in NGC~3379 and 50 PNe in NGC~3384;
NGC~4636 (2~PNe) and NGC~1549 (6~PNe) are included for completeness.
We present our samples in tabular form, as well as the spectrum for
each PN.  We constructed simple non-parametric spherical mass models
for NGC~3379 using a Monte Carlo Markov Chain method to explore the
space of likely mass models.  We find a remarkably constant
mass-to-light ratio within five half-light radii with an overall $B$
band mass-to-light ratio $\sim 5$.  A simple mass-to-light estimate
for NGC~3384 yields $\Upsilon_B \sim 11$, but is likely an
overestimate.
\end{abstract}


\section{Introduction}

The existence of Dark Matter (DM) has been postulated to resolve the
conflict between dynamical estimates of galaxy and galaxy cluster
masses, and estimates of their luminous or baryon mass.  DM, as a
source of gravitational potential, is an essential ingredient in the
current $\Lambda$CDM concordance cosmology of the Universe as derived
from observations of the Cosmic Microwave Background \citep{sperg+03}
and high redshift Supernova searches \citep{ries+98,perl+99}.  Little
consensus exists, however, on its properties at smaller galactic scales
with observations \citep[e.g.][]{debl+03} and numerical simulations
\citep[e.g.,][]{diem+04} currently being at odds.  Indeed, the lack of a
consistent understanding of DM has led some to question the necessity
of the DM postulate \citep{sk01}.

An important diagnostic in the study of DM is its distribution
throughout a galaxy. The strongest constraints come from observations
of spiral galaxies, where neutral hydrogen gas can be used to detect
the influence of DM beyond the stellar disk \citep{sv01}.  The case
for DM in the more massive elliptical galaxies is less concrete.
Unlike the simpler geometry of disk galaxies, the intrinsic
three-dimensional shape of an elliptical cannot be uniquely determined
from its projected shape.  Similarly, the orbital structure of
ellipticals is dominated by random motion, whereas the stars and gas
in spiral galaxies move on nearly circular orbits.  These fundamental
limitations are exacerbated by practical ones.  Because the galaxy
surface brightness drops steeply as a function of radius, measurements
of the stellar kinematics in ellipticals do not extend far into the
galaxy halo, and require substantial observational effort.
Furthermore, their gas content is low, and surveys of tracer
populations such as globular clusters and planetary nebulae have until
recently generated only modestly sized samples, the one exception
being Cen~A \citep{hea95}.  The picture is only discouraging in
contrast to the observations of disk galaxies, however, and---taken in
their own right---the recent data on elliptical galaxies have greatly
improved the understanding of these systems.

Measurements of the stellar kinematics in ellipticals now include
higher order moments of the line-of-sight velocity distribution
(LOSVD) as a matter of course \citep[e.g.,][]{caroea95}.  Knowledge of
the actual LOSVD, as opposed to only the first and second moments, can
lift the degeneracy between the mass distribution and the velocity
structure \citep{ger93,mer93}.  Along with more sophisticated modeling
techniques, the observations, sometimes going out as far as two
effective (half-light) radii $(R_e)$, have allowed for a better
constrained dynamical picture of a growing number of ellipticals
\citep[for references, see][]{kronea00}.  Most intriguingly, the
circular velocity curves in the \citet{kronea00} sample are quite
flat, indicative of the presence of a DM halo.  Despite this, no clear
trends have emerged about the DM distribution in ellipticals, with
some galaxies showing evidence for DM within $2R_e$ (mass-to-light
ratio in the $B$ band $\Upsilon_B \sim 20 \textrm{-}30$), and others
showing little or no evidence \citep{kronea00}.

Gas is scarce in early-type galaxies, but not rare.  Polar ring
galaxies, a rare breed of S0s with a ring of neutral hydrogen
perpendicular to the galaxy's symmetry plane, have been used to
constrain the mass and shape of DM halos, but much depends on the
assumption of self-gravity of the ring \citep{spar02}.  The most
consistent evidence for a DM halo around ellipticals comes from
studies of their hot gas, which radiates in the X-ray around 1~keV.
Using a sample of giant ($L>L_*$), ellipticals \citet{lw99} conclude
that DM does not dominate within an effective radius, but that the DM
fraction within $6R_e$ is in the range of 39-85\%.  Because early type
galaxies with $L<L_*$ have faint X-ray spectra which tend to be
dominated by the contribution from the stellar component, e.g. X-ray
binaries, this finding cannot be extended to fainter ellipticals.
Imaging by X-ray satellites \textit{Chandra} and \textit{XMM-Newton},
which shows bubbles and other structures, has complicated the picture
presented by \citet{lw99}, which assumes---as is typical---sphericity
and (isothermal) hydrostatic equilibrium.  For a recent review of the
topic of hot gas in ellipticals, see \citet{mb03}.

Gravitational lensing of light by elliptical galaxies is an additional
probe of DM halos, especially at large radii.  \citet{wilsea01} have
used the distortion (shear) of faint background galaxies by a
foreground galaxy to measure the mass-to-light ratio of early-type
galaxies.  Adding the shear signal from many hundreds of early type
lens galaxies in their fields they derive $\Upsilon_B \sim 121\pm28$
within $\sim 100\mr{\ kpc}$.  Strong lensing, where the lens galaxy
produces multiple---typically two or four---images of a background
source, yields more specific information, but the sample of lens
systems is limited $(<50)$.  \citet{keet01} deduces that DM cannot
account for more than 33\% of the mass within one $R_e$, or 40\%
within $2R_e$, consistent with the findings of \citet{lw99}.

Finally, globular clusters (GCs), planetary nebulae (PNe), and
satellite galaxies provide discrete tracers of a galaxy's kinematics
and, with suitable assumptions, of the dynamics \citep[e.g.,][]{rk01}.
GCs are not as numerous as PNe, and the additional difficulty of
absorption line spectroscopy on faint objects has limited the use of
GCs as kinematic tracers \citep[e.g.,][]{bridea03,cotea03}.  The
analysis is further complicated by the fact that GCs do not generally
trace the light distribution of their host galaxy.  In this paper we
focus on the Planetary Nebulae (PNe) that can be used to investigate
the dynamics of early-type galaxies.  This approach stretches back to
\citet{nf86} and their study of M32.  PNe are part of the brief phase
in the life of an intermediate mass (0.8--8~$M_\odot$) star when it
evolves from a red giant to a white dwarf; this makes PNe a good
tracer of old stellar populations (but see \citet{pengea05} for a
discussion on ``young'' PNe).  During this transition the star ejects
most of its mass, leaving a hot stellar remnant which rapidly
($\sim10^4$ years) cools to a white dwarf.  The PN proper is an
expanding shell of dense gas being photo-ionized by the central
stellar remnant.  PNe are bright objects, emitting on the order of a
few hundred $L_\odot$ in a few emission lines, most notably
[\ion{O}{3}] $\lambda5007$ \citep[up to 15\% of the total
luminosity;][]{djv92}.  This not only enables us to detect them at
extragalactic distances (out to 10--15~Mpc on 4-m class telescopes),
but also to measure their velocity along the line of sight.

We have collected PNe samples for a set of early-type galaxies with
the Rutgers Fabry-P\'erot (RFP), which can be thought of as a narrow
band ($\sim 2~\textrm{\AA}$) filter with a tunable central
wavelength.  Scanning around the (redshifted) [\ion{O}{3}]
$\lambda5007$ emission line, the RFP generates a three-dimensional
data cube where the galaxy background has been strongly reduced and
monochromatic point sources, such as PNe, stand out against the
background.  Detection and measurement of the radial velocity are all
achieved in one observing run.  Other techniques that look for
extragalactic PNe include the Planetary Nebula Spectrograph
\citep{dea02}, an instrument using slitless spectroscopy, and on/off
band photometry in conjunction with multi-object spectroscopy
\citep{pengea04,menea01}.

The paper is outlined as follows: \S\ref{s:observations} summarizes
the observing strategy and conditions; \S\ref{s:reduction} and
\S\ref{s:calibration} discuss the data reduction and calibration;
\S\ref{s:extraction} to \S\ref{s:compcont} show the process of our PN
candidate selection, including accounting for completeness and
interlopers; and we present the samples in full in \S\ref{s:results}.
Considering the recent interest in NGC~3379 \citep{romea03} we present
our mass models for this galaxy in \S\ref{s:massmodels}.  A summary of
our findings is given in \S\ref{s:wrapup}.

\section{Observations}
\label{s:observations}

We observed the four galaxies in our sample with the RFP over the
course of two runs at the CTIO 4~m Blanco telescope (f/8 Cassegrain
focus).  Details of the setup are given in Table~\ref{t:setup}.  The
RFP has a circular field of view with a diameter of $2\farcm8$, and a
spectral response function that is well approximated by a Voigt
profile with a $\mathrm{FWHM} \simeq 2$~\AA, equivalent to $120\
\mathrm{km}\ \mathrm{s}^{-1}$ at 5000~\AA.  There is a wavelength
gradient between the center and the edge of an image of 4.6~\AA.  We
typically scanned a wavelength span of 25~\AA\ around the
appropriately redshifted 5007~\AA\ [\ion{O}{3}] emission line in steps
of approximately 1~\AA.  At this wavelength the free spectral range of
the RFP is 48~\AA\ and we used one of two blocking filters with a FWHM
of 44~\AA\ to ensure that only one order was transmitted.  We used Tek
$1024\times1024$ pixel CCDs, with $0\farcs35$ pixels (binned to
$0\farcs70$ on the 1994 run), but only read out the portion of the CCD
illuminated by the RFP.

At the start of each run we observed a set of emission lines around
5000~\AA\ using a calibration lamp.  The resulting ``ring'' images
allowed us to establish the relationship between the gap setting $z$
and the transmitted wavelength $\lambda_c$ at the center of the image;
to determine the free spectral range of the etalon; and to measure the
shape of the spectral response function.  To monitor changes in the
wavelength calibration we took further images of the lamp throughout
each night.  We discuss the wavelength calibration in more detail in
\S\ref{s:wlcal}.  For each run we obtained one series of dome flats
at settings spaced $\sim1\ \textrm{\AA}$ apart.  Twilight flats are
not advisable, because of the spectral structure of the sky at the
resolution of the RFP.  Additionally we observed spectrophotometric
standard stars on each run: six 90~s exposures of LTT~3218 in 1994 and
six 300~s exposures of LTT~2415 in 1995 \citep{sb83}.

The Leo galaxies, NGC~3379 and NGC~3384, were observed with multiple
pointings, typically one pointing per night.  NGC~1549 and NGC~4636,
on the other hand, had only one pointing each, but were observed
throughout the course of a run to get the full wavelength coverage.
NGC~1549 was observed at the start of each night, NGC~4636 towards the
end.  The exposure time for each image was 900~s.  To avoid systematic
effects in the photometry consecutive exposures were not sequential in
wavelength and were dithered by a few pixels.  The appropriate
blocking filter was determined by the central wavelength in an
exposure being smaller or larger than the switching wavelength
$\lambda_s$ (see Table~\ref{t:setup}).  Exposures were repeated if the
seeing or the photometric conditions were particularly poor.  The
observations are summarized in Table~\ref{t:fields}.

The seeing was poor during both runs: $\sim1.5\arcsec$ in 1995 and in
the range 1.7--2.2\arcsec\ in 1994.  The latter is partly due to
difficulties in focusing the telescope and problems with the
auto-guider.  Conditions were mostly photometric in 1994, with the
exception of the last half of the first night, which only affected the
observations of NGC~4636.  For the 1995 run we were less fortunate.
The first two nights are fully photometric, but not the two following
nights.  As a result only half of our fields was observed completely
under photometric conditions.  As we will show below, this affected
the quality of our PN photometry, but not of our radial velocity
measurements.

\section{Data Reduction}
\label{s:reduction}

We used IRAF\footnote{IRAF is distributed by the National Optical
Astronomy Observatories, which are operated by the Association of
Universities for Research in Astronomy, Inc., under cooperative
agreement with the National Science Foundation.} for most of our data
reduction.  Each image was overscan corrected, trimmed and bias
subtracted.  Since the dark current contribution was negligible for
the length of our exposures, no dark frames were taken.  Each image
was flatfielded using the flatfield image with a central wavelength
closest to that of the image.

To aid the cosmic ray removal process and the photometry (see
\S\ref{s:flcal}) we aligned the images for each pointing (a ``stack'')
using the few stars that were visible in each field (see
Table~\ref{t:fields}).  In all cases a simple shift over a few pixels
proved to be adequate.  Since the PSF was well resolved, we shifted
the images by integer pixels to avoid interpolation effects.  Cosmic
rays were tagged, not removed, by combining all the images in a stack
using the IRAF task \texttt{imcombine} with the \texttt{crreject}
rejection algorithm, creating a mask for each image which covered the
comic ray events.  To be conservative we added a rim of masked pixels
around each event.  The padded masks covered on average 8\% of the
field of view.

The final step in preparing the images in a stack was the removal of
the background light.  In our case the background is mainly light from
the galaxy and its ghost, with the sky contributing an almost
negligible amount.  Ghosts are the result of the reflections between
the CCD and the etalon, which create a faint in-focus replica of the
original object, mirror reflected about the optical axis.  We removed
the background to reduce gradients which would bias the photometry,
allowing us to detect PNe closer to the galaxy.  We wrote a Fortran
program which estimated the background with a ring filter
\citep{sec95}, where we replaced the median with a more robust
estimator of the mean \citep{hmt83}.  A ring filter removes objects at
scales smaller than the diameter of the ring and we chose our diameter
to be about twice the seeing radius.  The shifted, masked and
background subtracted stacks formed the basis for our further
analysis.

\section{Calibration}
\label{s:calibration}
The calibration of the RFP stacks consists of three parts: (i) image
registration, i.e. assigning celestial coordinates to each pixel
position; (ii) spectral calibration; (iii) flux normalization,
i.e. accounting for variations in the observing conditions.

\subsection{Image Registration}
\label{s:imgreg}

In each field we had only a few stars to register the images (see
Table~\ref{t:fields}).  These astrometric reference stars generally
did not have published positions, because of their proximity to the
galaxy being observed, and we had to determine their coordinates from
observations with a larger field of view.  For NGC~3379 and NGC~3384
we had deep broad band ($V$-like) images with $0\farcs4$ pixels; for
NGC~1549 and NGC~4636 we used images from the Digitized Sky
Survey\footnote{The Digitized Sky Survey was produced at the Space
Telescope Science Institute under U.S. Government grant NAG
W-2166. The images of these surveys are based on photographic data
obtained using the Oschin Schmidt Telescope on Palomar Mountain and
the UK Schmidt Telescope. The plates were processed into the present
compressed digital form with the permission of these institutions.}
(DSS) with $1\farcs7$ pixels.  Using the USNO-A2.0 catalog
\citep{mea98} we were able to identify 80 or more stars in each of
these images and consequently measured the celestial coordinates of
the RFP reference stars.

The systematic uncertainty in a single RFP coordinate, as estimated
from the variance in the reference star positions, is $0\farcs6$ for
NGC~1549 and NGC~4636, compared to about $0\farcs1$ for the other two
galaxies.  In the case of the Leo~I galaxies (NGC~3379 and NGC~3384)
we noticed a systematic $\sim 1''$ offset between the \citet{cjf89}
coordinates and our own when we compared our PNe sample with their
lists (see \S\ref{s:samp3379}).  We found that the offset can be
attributed completely to the difference in astrometric reference star
coordinates.  Since our project does not require absolute astrometry
we did not investigate the matter further.

\subsection{Spectral Calibration}
\label{s:wlcal}

The transmitted wavelength $\lambda$ at a position $\bm{x}=(x,y)$ on
an RFP image taken at a gap setting $z$ is given by
\begin{equation}
   \lambda(x,y,z) = (a+bz) (1+ |\bm{x}-\bm{x}_c|^2 / f^2)^{-1/2},
   \label{e:wlgrad}
\end{equation}
where $a$ and $b$ are parameters relating $z$ to the central
wavelength of an RFP image, $\bm{x}_c$ is the position of the optical
axis of the RFP and $f$ is the focal length of the camera lens of the
RFP.  To calibrate this relation we observe several emission lines
from a calibration lamp at a range of $z$ settings at the beginning of
a run.  During a run we take additional calibration images (``night
rings''), eight to ten per night, to ensure the stability of our
calibration.  Over the course of our runs $b$ and $f$ were stable,
consistent with our experience from other runs.  The value for
$\bm{x}_c$, however, fluctuates as a result of flexure in the
spectrograph; likewise, the wavelength zero-point $a$ shifts due to
drift in the control electronics.  The overall drift in $a$ was a
little larger than 1~\AA\ over a run; the overall shift in $\bm{x}_c$
never larger than 6~pixels during a night.  We interpolated the values
for $a$ and $\bm{x}_c$ for each image from the values obtained from
the night rings.  The uncertainty in $\lambda$ due to the calibration
is 0.05~\AA\ ($3\ \textrm{km}\ \textrm{s}^{-1}$ at 5007~\AA), based
on the scatter in our calibration fits with the dominant source of
uncertainty being $f$.

The initial calibration run also allows us to calibrate the spectral
response function (SRF) of the RFP, i.e. the line shape of a
monochromatic source.  Although for a perfect etalon the SRF is an
Airy function, experience shows that the SRF is better fitted by a
Voigt function, the convolution of a Gaussian with a Lorentzian
\citep{vhr47}.  The shape is determined by two parameters: the
Gaussian width $\Delta\lambda_G$ and the Lorentzian width
$\Delta\lambda_L$ (see Table~\ref{t:setup}).  Note that the shape
parameters are significantly different between the two runs, but that
the full width at half maximum (FWHM) of the profile is practically
the same at $\sim2$~\AA.  We intentionally did not want to resolve the
intrinsic structure of the PN line profile (see
Section~\ref{s:modspectra}), since it would reduce the depth of our
survey and would complicate our study unnecessarily.

\subsection{Flux Normalization}
\label{s:flcal}

In each field of view we use a reference object to provide us with a
fiducial flux for every image in a stack, to calibrate the extinction
due to light cirrus.  Considering the narrow wavelength range we are
observing it is reasonable to assume a flat spectrum for each
reference object.  We can then calculate normalization factors for
each image in a straightforward manner.  Only the spectrum of the
reference star in the West field of NGC~3379 showed an indication of
absorption lines.  For NGC~3379 we therefore decided to use the galaxy
flux within a $10\arcsec$ aperture as a reference source.  The galaxy
spectrum has a slight curvature at the blue edge, which we modeled
with a smoothing spline.  The uncertainty in each normalization factor
is 3-5\%.

We were unable to get an independent determination of the atmospheric
extinction during our runs, because we observed our spectrophotometric
standard stars only once a night and all at approximately the same
airmass.  Instead, we adopted a value for the extinction of
0.19~magnitude per airmass as given for CTIO \citep{hea92}; on
photometric nights the flux normalization factors were found to be
consistent with assumed extinction.  Finally, using the
spectrophotometry for the standard stars, we converted our
instrumental fluxes to physical units.  Comparison with previously
published PN magnitudes shows no systematic effects (see below).

\section{Extracting Spectra}
\label{s:extraction}

We extract spectra for every independent point in the field of view of
a stack and for each spectrum decide whether or not an emission line
is present.  A more targeted approach, where we would look for
emission line point sources in each image of a stack using, for
instance, DAOPHOT \citep{pbs87}, requires extensive fine tuning due to
the low signal-to-noise ratio of the objects and still produces
numerous spurious detections.  (\citet{tmw95} used this approach to
look for PNe around NGC~3384.)  Our method is not computationally
expensive and has no bias due to a search algorithm.

The fluxes for each spectrum were extracted using the \texttt{apphot}
aperture photometry package in IRAF.  We performed a set of Monte
Carlo simulations of our instrumental setup (see Table~\ref{t:setup})
to compare the results of PSF and aperture photometry.  PSF
photometry, assuming perfect knowledge of the PSF, produced an
appreciably larger scatter in the derived fluxes than aperture
photometry did.  Because we have only one star in each field bright
enough to determine a PSF, there was the additional concern of the
effects of a PSF mismatch.  Both considerations lead us to use
aperture photometry.

Each field of view was sampled on a triangular grid, the most
isotropic choice.  The spacing between sample points was chosen to
Nyquist sample the image, i.e. sample points are separated by half the
smallest seeing FWHM in a stack, to guarantee no information was lost.
Following \citet{nay98} we chose our aperture radius to be slightly
larger than two thirds of the seeing FWHM in order optimize the signal
to noise ratio within the aperture.  In this case the photometry
apertures overlap slightly more than 50\% in area.  If the aperture
contained masked pixels, e.g. a cosmic ray, the data point was removed
from the spectrum.  In cases where we measured a negative flux, we
estimated the uncertainty from the uncertainties in the positive
fluxes near the continuum level.  No aperture corrections are needed,
since we photometer a PN candidate and a reference star in the same
way.

The final data product, then, is a set of $M$ spectra where each
spectrum is a list of the form $\{\lambda_i, f_i, \sigma_{f_i},
e_i\}_{i=1}^N$ with $N$ the number of images in a stack, $\lambda_i$
the wavelength, $f_i$ the flux, $\sigma_{f_i}$ the uncertainty in the
flux and $e_i$ a possible photometric error flag.  In the following
section we discuss how we determine which spectra indicate the
presence of an emission line.

\section{Candidate Selection}
\label{s:selection}

To establish the presence of a PN candidate in a spectrum we fit a
flat continuum model and an emission line model to each spectrum,
measuring the difference in the goodness-of-fit using a statistic $S$.
After calibrating the distribution of $S$ with Monte Carlo simulations
we choose a significance level $\alpha$ and select all spectra with a
value $S>S(\alpha)$.  Finally, we visually inspect each stack at the
candidate positions to verify the selection and avoid contamination
from CRs and image artifacts.  The details of the procedure are
described below.

\subsection{Modeling the spectra}
\label{s:modspectra}
We expect most extracted spectra to have a flat, practically zero,
flux distribution, since we removed the background in each image.  In
addition a flat model spectrum should encompass the spectra from
foreground stars in our field.  Hence, our null hypothesis is a simple
constant flux model.  Modeling spectra with a linear function $f_i =
a\lambda_i+b$ yielded no significant improvements over the constant
model, because of the flux uncertainties.

The alternative hypothesis is that a spectrum contains an emission
line.  Such a model needs to take into account the intrinsic structure
of the [\ion{O}{3}] line, since PNe, as observed in our own Galaxy,
often have doubly peaked emission lines \citep{pot84}.  The intrinsic
width of each of these peaks is small ($\sim10\ \textrm{km}\
\textrm{s}^{-1}$), and the nebular expansion velocity $v_\mr{exp}$,
defined as half the separation between the two peaks, has a
distribution with a mode around $10\ \textrm{km}\ \textrm{s}^{-1}$,
but a mean of $25\ \mr{km}\ \mr{s}^{-1}$ due to the large tail at
higher velocities \citep{phi02}.  Based on the latter we calculate
that the broadening of our \emph{observed} line profile (see below)
due to line structure is $\lesssim9\%$, negligible considering the
wavelength sampling of our spectra (see Table~\ref{t:setup}).  We
will assume from now on that PNe can be treated monochromatic point
sources.

Consequently, the line profile is determined by the effective filter
transmission $T$ and the RFP spectral response $R_s$.  We separate the
two because they are measured separately from each other.  The
measurement of the latter is described in section~\ref{s:wlcal}; for
the former we use the filter curves provided by CTIO.  We assumed that
temperature effects on $T$ are unimportant.  The f/7.5 beam used for
our observations is slow enough to have a negligible effect on the
transmission properties of our filters \citep{cjf89}.

The effective filter transmission $T$ changes the line shape from a
simple Voigt profile in two ways.  The first modification is a
discontinuous jump in the emission line profile, which is the result
of changing from the ``blue'' filter $T_b$ to ``red'' filter $T_r$
whenever the central wavelength of an image is larger than the
switching wavelength $\lambda_s$.  The total flux of a PN observed at
$\lambda_\mr{pn}$ will differ between the two filters, since in
general $T_b(\lambda_\mr{pn}) \neq T_r(\lambda_\mr{pn})$.  Hence, a
spectrum will show a jump at $\lambda_\mr{s}$.  The jump is
appreciable when $\lambda_\mr{pn}$ close to $\lambda_\mr{s}$, but is
otherwise negligible.

\begin{figure}
\epsscale{0.65}
\plotone{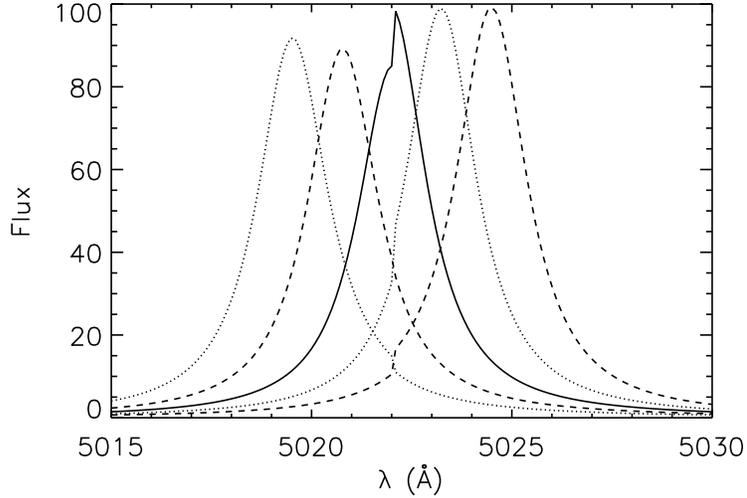}

\caption{Line profiles of five PNe of equal brightness with
wavelengths 1.25~\AA\ apart.  The solid line is a PN at a wavelength
of 5022~\AA, which is also the switching wavelength.  The line
profiles take into account the effects of filter switching (jump) and
flatfielding (luminosity bias).  The flux units are arbitrary.
Different line styles were used for clarity.}

\label{f:profiles}
\end{figure}

The second modification stems from the fact that the flatfield images
are based on exposures of continuum sources.  A flatfield image taken
on the red edge of a filter includes light from the neighboring order
of the RFP on the blue edge.  As a result we overestimate the
sensitivity of the detector when applying the flatfield to an emission
line object, which has flux in one order only.  Ignoring this effect
would lead to a wavelength dependent bias in the measured total flux
of a PN.  We can adjust for this effect by calculating a correction
factor $\Gamma$, the integral over the filter transmittance and the
SRF.  Figure~\ref{f:profiles} illustrates how the two effects modify
the line profile.
 
To account for the above two effects we characterize the spectrum of a
PN candidate extracted from a stack of $N$ images in the following way:
\begin{equation}
   \label{e:emmodel}
   M(f_\mr{pn},
   \lambda_\mr{pn}, c_\mr{pn}|\bm{x}, z_i) =  f_\mr{pn} {
   R_s(\lambda_\mr{pn}, \bm{x},z_i)
   T(\lambda_\mr{pn},z_i) \over \Gamma(\bm{x},z_i) } +  c_\mr{pn},
   i=1,\cdots,N, 
\end{equation}
where $M$ is the flatfielded and normalized flux measured at position
$\bm{x}$ and RFP setting $z_i$, $f_\mr{pn}$ is the total flux received
from the PN candidate, $R_s$ is the SRF, $T(\lambda_\mr{pn})$ is the
appropriate filter transmission, and $c_\mr{pn}$ is the continuum
level.  Since the spectra we extract from the RFP stacks are
critically sampled, we cannot constrain the shape of the line profile
in addition to fitting a peak wavelength, total intensity and
background flux.  Hence, we fixed $\Delta\lambda_G$ and
$\Delta\lambda_L$ to the values determined from the calibration run in
our further data analysis.  The factor $\Gamma(\bm{x},z_i)$ corrects
for the luminosity bias; it depends implicitly on $R_s$ and $T$.  Note
that $c_\mr{pn}$ can potentially give us a handle on the contamination
of our sample (see section~\ref{s:cont}).

We fit the two models to each spectrum by minimizing a goodness-of-fit
statistic $s^2(f_\mr{pn},\lambda_\mr{pn}, c_\mr{pn})$ defined by 
\begin{equation}
   s^2 = \sum_{i=1}^N {[f(\bm{x}, z_i) - M(f_\mr{pn},
   \lambda_\mr{pn}, c_\mr{pn}|\bm{x}, z_i)]^2 \over \sigma_{f_i}^2 }.
\end{equation}
We set $f_\mr{pn} \equiv 0$ for the continuum model and $f_\mr{pn}
\geqslant 0$ for the emission line model.  The minimum $s^2$ value was
found using the E04UNF routine from the NAG numerical library, which
is designed to solve nonlinear least-squares programming problems in
the presence of constraints on the parameters.  Since the
uncertainties in our data points are not gaussian, the uncertainties
in our best fit parameter values do not necessarily correspond to a
$1\sigma$ error.  We used simulated observations (see
Section~\ref{s:finsel}) to verify the plausibility of uncertainties
found in the fitting procedure.

\subsection{Choosing the correct model}

The continuum model $M_c$ and the emission line model $M_l$ are nested
models, i.e. the parameters in $M_c$ form a subset of the parameters
in $M_l$ (the additional parameters of $M_l$ are fixed to some default
value).  The likelihood-ratio (LR) test or the $F$-test are
conventionally used to decide between the null hypothesis $M_c$ and
the alternative hypothesis $M_l$ \citep{eea71,bea97}.  Each of these
tests uses a test statistic $S$ which is some simple function of $\hat
s^2_c$ and $\hat s^2_l$, the best fit values of $s^2$ for $M_c$ and
$M_l$, respectively.  Under certain regularity conditions these test
statistics have analytically known reference distributions
(e.g. \citealt{pea02}), usually in the limit of large $N$, which allow
a level of significance to be determined for the observed value of
$S$.

In our case two of the regularity conditions are not met.  First, the
tests assume that the data are independent, identically distributed
random variables.  As a result of the flatfielding the data points in
our spectra are not identically distributed.  Second, the default
values for the additional parameters in $M_l$ cannot be on the
boundary of parameter space.  The null hypothesis assumes that
$f_\mr{pn} \equiv 0$, clearly on the boundary of our parameter space.
This is not an academic point: \citet{mea96} show that when this
condition is violated the actual reference distribution is markedly
different from the nominal reference distribution (see their
Figure~3).  The problem of including boundary values in a parameter
space has no standard analytical solution.  \citet{pea02} present a
more comprehensive introduction.

We decided to use the LR test statistic $S = \hat s^2_c - \hat s^2_l$,
but to calibrate its reference distribution by way of Monte Carlo
simulations instead of using the nominal reference distribution.  (The
choice of the LR test over the $F$-test is motivated in the next
section.)  The reference distribution $p(S|M_c) \mr{d} S$ specifies
the probability that the test statistic has an observed value $S$
under the condition that the null hypothesis is true.  To calibrate
$p(S|M_c)$ we simulated stacks of images that contain no emission line
objects, preserve the noise characteristics of the original
observations, and are reduced in the same way as the original
observations.  Every field of view was calibrated separately.

\begin{figure}
\epsscale{0.65}
\plotone{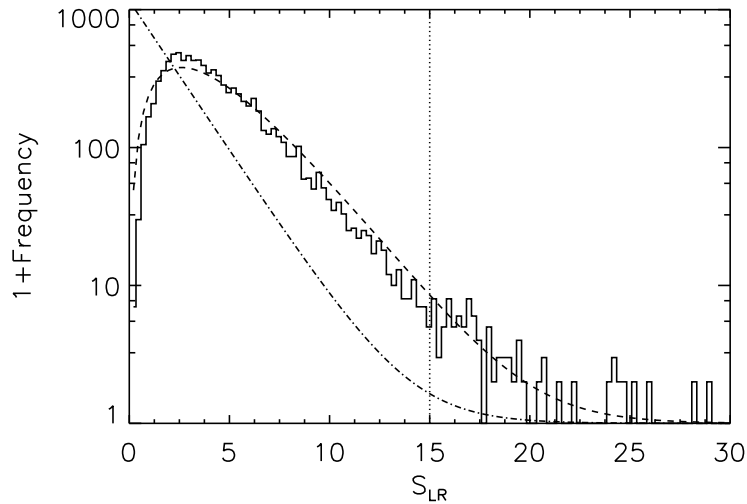}

\caption{Histogram of the $S_\mr{LR}$ values from a simulated set of
blank stacks based on the observations of the NGC~3379 East field.
Note the logarithmic scale of the vertical axis.  The vertical dotted
line indicates the 1\% upper tail of the distribution.  The dot-dashed
line is a $\chi_2^2$ distribution, normalized to our sample size.  The
dashed line is a $\chi_{4.6}^2$ distribution, similarly normalized,
plotted for comparison.}

\label{f:falsepos_hist}
\end{figure}

As an example, we show the result of these simulations for the East
field of NGC~3379 in Figure~\ref{f:falsepos_hist}.  The nominal
reference distribution of $S$ in our case is a $\chi_2^2$ distribution
\citep{eea71}.  In this particular case the actual reference
distribution can be approximated by a $\chi^2_\nu$ distribution, as
shown by the best-fitting $\chi_{4.6}^2$ curve, but this is not
generally true.  A proper calibration of the test statistic is clearly
essential in order to avoid a plethora of false positives.  For each
field of view we choose the value of $S_c(\alpha)$ that corresponds to
a significance level $\alpha = 0.01$ and select all spectra with
$S>S_c(\alpha)$.  Our experience shows that at this significance level
we can observe a PN candidate in at least two frames.

\subsection{Final selection}
\label{s:finsel}

Since the photometric apertures overlap, PNe candidates in each field
of view tend to cluster in contiguous groups of 3-6 spectra sharing a
similar central wavelength $\lambda_\mr{pn}$.  Instead of developing
an algorithm to make a final selection of candidates from these
clusters, we performed the final selection by hand.  First we remove
any obviously false identifications due to CRs and image artifacts,
e.g. the ghost image of a bright star. Next we choose the brightest
spectrum in a cluster as the candidate spectrum and visually inspect
its position in every image of the stack.  Once we are convinced we
have a bona fide PNe candidate, we determine the exact position of the
candidate and redo the photometry of the candidate to provide the
final parameters for the object.

We have two estimates for the uncertainties in $f_\mr{pn}$,
$\lambda_\mr{pn}$, and $c_\mr{pn}$: the variances from the formal
best-fit covariance matrix \citep{eea71}, and the empirical estimates
from the artificial PNe simulations (see section~\ref{s:compcont}).
For PN candidates with $S>S_c(\alpha)$ we found that the best-fit
parameters are only slightly correlated and that the two uncertainty
estimates are consistent with each other.  Hence, we used the best-fit
variances as our measure of the uncertainties in the magnitude and
line-of-sight velocity.  We note that $\lambda_\mr{pn}$ is in our
experience significantly better constrained than $f_\mr{pn}$; the
continuum level $c_\mr{pn}$ is always consistent with zero, as
expected.  The false-negative simulations also allow us to estimate
the random uncertainty in a PN position at $0\farcs3$, which does not
include the systematic uncertainty discussed in
section~\ref{s:imgreg}.

In the same way that one can determine an aperture radius that
optimizes the signal-to-noise ratio $Q$ within some photometric
aperture \citep{nay98}, we determine a wavelength window centered on
the peak wavelength which optimizes $Q$ in the spectral domain.
Numerical experiments show that a window width of 4/3 the spectral
FWHM is optimal.  To calculate the signal-to-noise ratio we simply
added the signal from all frames within 1.4~\AA\ of $\lambda_\mr{pn}$
and divided this by the sum in quadrature of the uncertainties (all
quantities were converted to photons).  Taking into account the
seeing, the background flux, the sampling and the photometric
conditions of our observations, we find that our empirical values are
in agreement with the theoretical expression of
\citet[eq.~6]{dea02}.

\section{Completeness and Contamination}
\label{s:compcont}

To characterize our PN samples fully we need to address two issues:
completeness and contamination.  The former describes the probability
that our experiment, i.e. the whole of observations, data reduction
and analysis, would not detect a PN with a flux $f_\mr{pn}$.  The
latter specifies the extent to which candidates in our samples are not
actually PNe.  In the context of our experiment the completeness is
closely related to the (statistical) power of the test statistic and
we show that the LRT is preferred over the $F$-test because it gives
us a more complete sample.  It is more difficult to quantify the
contamination, mainly because the distribution of possible
contaminating sources is still poorly understood.  Based on our
estimates, discussed in more detail below, contamination is not a
serious concern in our samples.

\subsection{Completeness and power}

For statisticians the usefulness or \textit{power} of a test lies in
its ability to reject the null hypothesis ($f_\mr{pn}=0$) when in fact
the alternative hypothesis ($f_\mr{pn}>0)$ is true.\footnote{For
purposes of detection, $\lambda_\mr{pn}$ and $c_\mr{pn}$ are
parameters of little interest (``nuisance parameters'') and we
marginalize the distribution of $S$ over these two parameters.}  The
power function  $\beta(f_\mr{pn})$ is quantified by 
\begin{equation}
   \beta(f_\mr{pn}) = p(S<S_c(\alpha)|f_\mr{pn}),
\end{equation}
i.e. the probability that, given a level of significance $\alpha$, we
reject the true (alternative) hypothesis \citep{eea71}.  The actual form of
$\beta(f_\mr{pn})$ is contingent upon our choice of $\alpha$, but is
generally monotonically increasing.  When multiple tests are
available, statisticians look for the test with a minimum
$\beta(f_\mr{pn})$, the one least likely to lead to an invalid
conclusion.

Astronomers, on the other hand, think in terms of completeness: we
want to maximize our ability to identify objects of a given type.
Since our experiments are limited by observing conditions (exposure
time, seeing, etc.) we rarely obtain a complete sample, instead
quantifying the level of completeness in terms of a limiting
magnitude.  Typically, the depth of a survey is defined by the flux
level where the probability of detecting an object is 50\%:
$\beta(f_\mr{lim}) = 0.5$ \citep{har90}.  Hence, the more powerful
test is the test where $f_\mr{lim}$ is the smallest and so allows the
survey to go deepest.

We considered two statistics commonly used in model comparisons: the
LR statistic
\begin{equation}
   S_\mr{LR} \equiv \hat s^2_c - \hat s^2_l,
\end{equation}
and the $F$-statistic
\begin{equation}
   S_F \equiv { s^2_c - s^2_l \over s^2_l} \left/ {N-\Delta P \over
   P_l} \right.,
\end{equation}
where $N$ is the number of data points in a spectrum, $P_l=3$ the
number of parameters in $M_l$, and $\Delta P=2$ the difference in the
number of parameters between $M_c$ and $M_l$.  A third statistic that
is sometimes used, the Goodness-Of-Fit statistic $S_\mr{GOF} \equiv
\hat s^2_c$, was not considered, since it does not take into account
an alternative hypothesis and consequently is not a powerful test.
Although $S_\mr{LR}$ and $S_F$ yield about the same significance for a
spectral feature \citep{bea97}, the LR test is generally more powerful
than the $F$-test, the latter being the more appropriate test when the
uncertainties in the data are unknown \citep{fea99}.

\begin{figure}
\epsscale{0.7}
\plotone{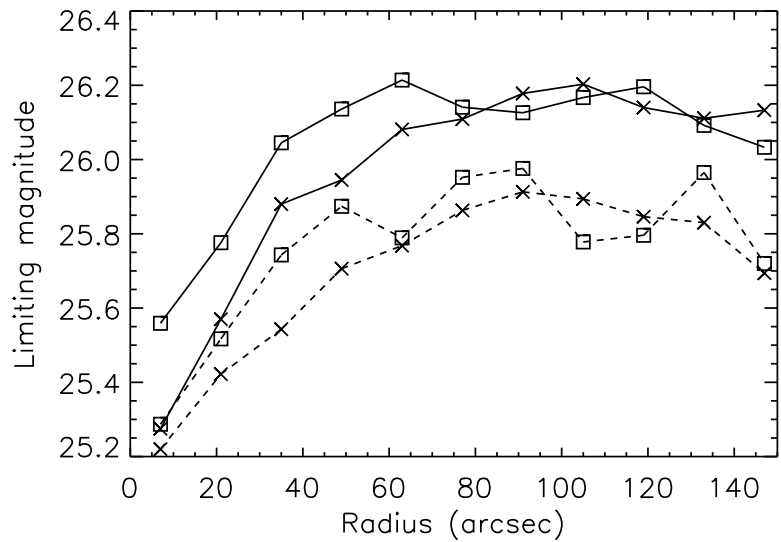}

\caption{Limiting magnitude as a function of the distance to the
center of NGC~3379.  The crosses and squares represent the East and
West field, respectively.  The solid lines are the limiting magnitudes
using the LR statistic; the dashed lines using the $F$-statistic. }

\label{f:3379maglim}
\end{figure}

In order to ascertain the limiting magnitude of our observations we
simulated stacks with artificial PNe and analyzed them as we would
real observations.  By comparing the analyses using $S_\mr{LR}$ and
$S_F$ we found that the former is the more powerful of the two tests,
as expected.  An illustration of this point is given in
Figure~\ref{f:3379maglim}: in the case of our NGC~3379 observations
$S_\mr{LR}$ allows us to go $\sim0.3$~mag deeper than $S_F$.  The
limiting magnitudes for each field of view are shown in
Table~\ref{t:fields}.

\subsection{Sources of contamination}
\label{s:cont}

Each PN detection is based on the assumption that the detected
emission line is the [\ion{O}{3}] 5007 line.  A simple way to
check if a candidate is actually a PN, short of taking a complete
spectrum, is to measure the flux in the [\ion{O}{3}] $\lambda4959$
emission line, since the line ratio $I(5007)/I(4959)=3$ is fixed
\citep{fea00}.  Our wavelength range, however, does not include the
redshifted $\lambda4959$ line, and we have to rely on statistics to
appraise the possibility of an interloper in our PN samples.

The search for high-redshift emission line galaxies in order to
determine their star-formation history \citep{hcm98,sea00}, as well as
searches targeted to quantify the contamination of intracluster PN
surveys \citep{kea00,cea02} have improved our understanding of
possible contaminants.  The most likely candidates that---to our
knowledge---could contaminate our samples are $\textrm{Ly}\alpha$
sources at $z = 3.13$, and [\ion{O}{2}] $\lambda3727$ at $z = 0.35$
\citep{kea00}.  In the absence of any other information, the most
likely interloper is a $\textrm{Ly}\alpha$ source, since these tend to
have significantly stronger lines than the [\ion{O}{2}] sources
\citep{cea02}.  In particular, at the depth of our observations
($m_{5007} < 26.6$ or $f_{5007} > 0.7\times10^{-16} \mr{\ erg\
cm^{-2}\ s^{-1}}$) the probability of the latter contaminating our
samples is negligible.  For the former \citet{cea02} determine a
surface density of $\sim 3500\ \mr{deg^{-2}}$ per unit redshift for
$\textrm{Ly}\alpha$ with $f_{5007} > 0.5\times10^{-16} \mr{\ erg\
cm^{-2}\ s^{-1}}$, consistent with the determination of \citet{cr03}
for the Leo group of galaxies.  Taking into account the redshift range
($\Delta z \sim 0.005$) and the effective area
(5.8--16.2~$\mr{arcmin}^{-2}$) of our surveys, we expect well fewer
than one contaminating source in any of our samples.

The estimate does not take into account any clustering of the
background sources anticipated because of large scale structure, which
could result in significant fluctuations in the surface density
\citep{oea03}.  In many cases, however, the $\mr{Ly}\alpha$ line will
be broad enough to be resolved by the RFP and have an asymmetric
profile \citep[e.g.,][]{cr03}.  Such sources are unlikely to be
selected in our detection procedure and would be recognizable from
their spectra.  Additionally, these $\mr{Ly}\alpha$ sources often have
a faint continuum which can be used to identify an interloper, which
partly motivated our inclusion of a continuum term in
equation~\ref{e:emmodel}.  All our candidate spectra show continuum
levels fully consistent with a zero background.  On the basis of these
considerations we believe that the possibility of contamination by
$\mr{Ly}\alpha$ sources in our samples can be ignored.

\section{Results}
\label{s:results}

\begin{figure}
\epsscale{1.0}
\plotone{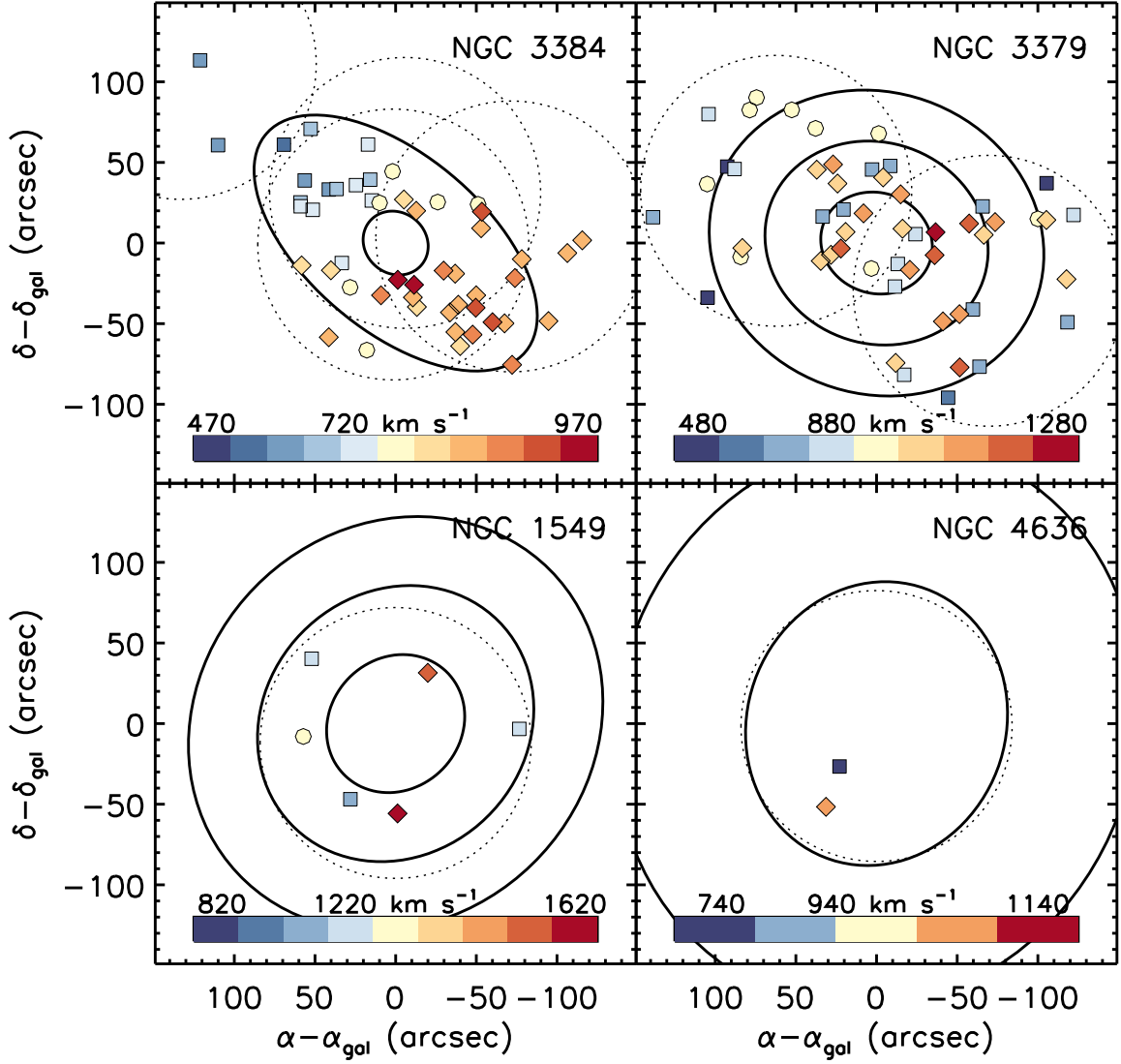}

\caption{Line-of-sight velocities of the PNe in our galaxy sample.
  Dotted lines indicate the various pointings for each galaxy (see
  Table~\ref{t:fields}); solid lines are the isophotes at $R_e$,
  $2R_e$, and $3R_e$, except for NGC~3384 with isophotes at $R_e$
  (bulge) and $\mu_B=25\mr{\ mag\ arcsec^{-2}}$ (disk).  Red diamonds
  (blue squares) indicate PNe moving at radial velocities greater
  (less) than the systemic velocity; buff circles show radial
  velocities close to the systemic velocity.  North is up and East is
  to the left.}

\label{f:rvfields}
\end{figure}

Our galaxy sample consists of three close to round elliptical galaxies
and one lenticular galaxy.  An overview of the properties of our
galaxy sample is given in Table~\ref{t:galsamp}.  We present the
spectra for each PN in our samples, as well as tables with positions,
magnitudes, and line-of-sight velocities.  A simple overview of the
line-of-sight velocities for all four galaxies is shown in
Figure~\ref{f:rvfields}.

\subsection{NGC~4636}

The E1 galaxy NGC~4636 lies on the Southern edge of the Virgo cluster
and has been well studied because of its high X-ray luminosity.
\citet{rea01} note that the galaxy does not follow the Fundamental
Plane relation for core galaxies (its surface brightness is too low
for its absolute luminosity) and has an unusually diffuse core.  This
is reflected in the large effective radius of the system.

All the NGC~4636 images were taken at the end of each night during the
1994 run.  Of the 34 available image 28 were centered on the galaxies,
but 6 were offset by half a field radius.  Additionally, the seeing of
these observations was large ($2\farcs2$) and the first night we had
non-photometric conditions.  Because of the lower surface brightness
we were able to go deep enough to discover two PNe in this system,
whose velocities are consistent with the systemic velocity
\citep{mce95}.  The spectra are shown in Figure~\ref{f:n1549spectra}.1
and the properties of the PNe in Table~\ref{t:1549-pne}.

\subsection{NGC~1549}

The E1 galaxy NGC~1549 is in close interaction with its neighbor
NGC~1553 as is evidenced by the strong isophote twisting \citep{fih89}
and the faint shells surrounding it \citep{mc83}.  The galaxy is the
most distant in our sample and has not been targeted for a PN survey
before.  Photometric conditions varied throughout the five nights of
observation, but this was mitigated by the relatively good seeing and
the small airmass of the observations, which allowed us to go a little
deeper than for the Leo~galaxies.  The galaxy was observed with one
pointing, giving an effective survey area of $5.8\mr{\ arcmin^{-2}}$.
We discovered 6 PNe around this galaxy.  The spectra are shown in
Figure~\ref{f:n1549spectra}.1 and the properties of the PNe in
Table~\ref{t:1549-pne}.  Their average velocity of $1279\mr{\ km\
s^{-1}}$ is consistent with the $1220\mr{\ km\ s^{-1}}$ systemic
velocity of the galaxy and their velocity dispersion $\sigma =
217\mr{\ km\ s^{-1}}$ matches the value found with absorption line
spectroscopy \citep[$220\mr{\ km\ s^{-1}}$,][]{lea94}.

\subsection{NGC~3379}
\label{s:samp3379}

\begin{figure}
\plotone{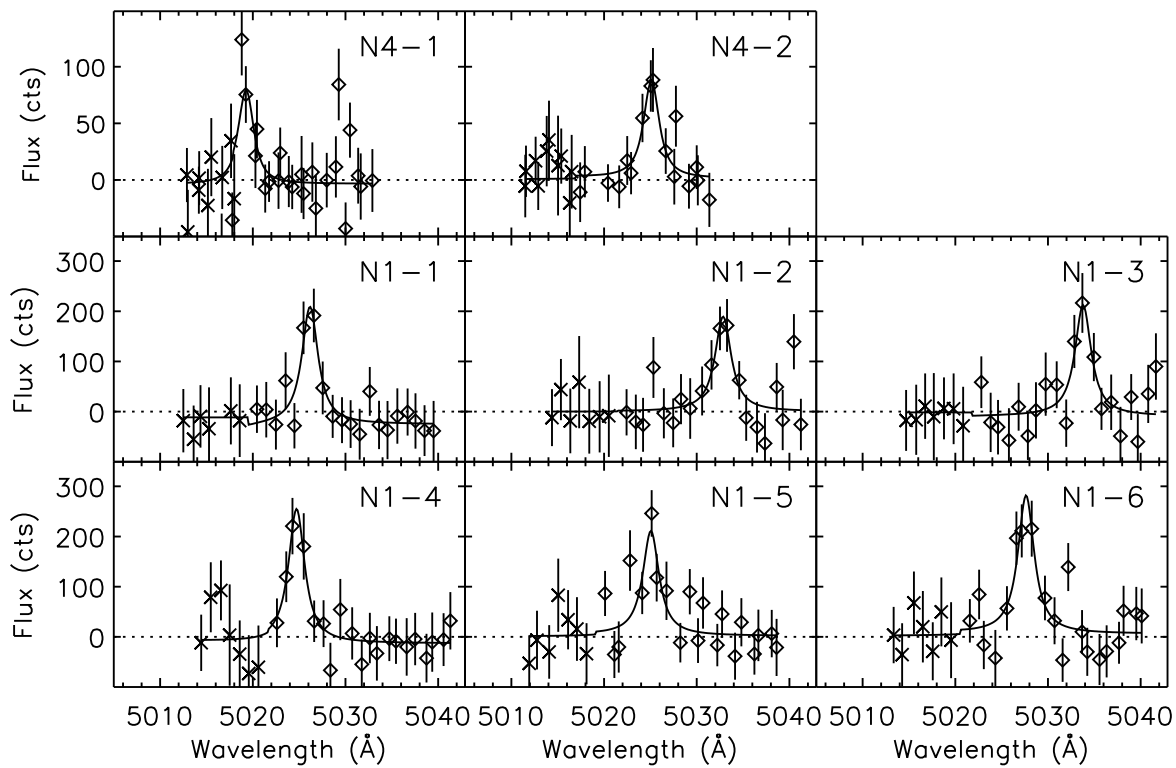}

\caption{RFP spectra of the NGC 1549 and NGC 4636 PNe.  Crosses denote
observations under the 5007-44 filter, squares the 5037-44 filter.
The curve is the best fit line profile as described by
equation~\ref{e:emmodel}.  The PN ID from Table~\ref{t:1549-pne} is
shown in the top right corner of each spectrum.  Note that the flux
scales between the two sets of PNe are not directly
comparable. Figures~\ref{f:n1549spectra}.1--\ref{f:n1549spectra}.7 are
available in the electronic edition of the Journal.  The printed
edition contains only a sample.}

\label{f:n1549spectra}
\end{figure}

The E1 galaxy NGC~3379 (M105) forms with NGC~3384 (below) the central
pair of the Leo (M96) group \citep{fs90}.  Following the work by
\citet{dvc79} NGC~3379 is often considered the ``standard''
elliptical, although \citet{cea91} have argued, purely on the basis of
photometry, for a reclassification from E1 to S0.  Because of its
proximity NGC~3379 is an ideal candidate to look for extragalactic PNe
and we observed it with two pointings, giving an effective survey area
of $11.9\mr{\ arcmin^{-2}}$, which excludes the bright central
($<10\arcsec$) part of the galaxy.  The frames for each pointing were
taken over two nights; each pointing had one night of non-photometric
quality.  Because one of our reference stars showed evidence of an
absorption line in its RFP spectrum, we used the galaxy flux within
$0.5R_e$ as the photometric reference for our flux normalization.  We
present our sample of PNe in Table~\ref{t:3379-pne}; the corresponding
spectra are shown as
Figures~\ref{f:n1549spectra}.2--\ref{f:n1549spectra}.4 in the online
version of the Journal.

In NGC~3379 we find 54~PNe out to $\sim5R_e$.  The sample is sparse
and its distribution on the sky is consistent with the surface
brightness of NGC~3379, taking into account the position dependent
limiting magnitude.  Using the on/off band technique \citet{cjf89}
obtained a sample of 93 PNe in NGC~3379.  In a spectroscopic follow up
\citet{cjd93} measured the line-of-sight velocities of a subset of 29
PNe.  In principle we can observe 57 of the \citet{cjf89} candidates,
and we recovered 30 of these.  Similarly, we can measure 15 velocities
from \citet{cjd93} and we recover 7 of these.  The overlap with the
previous work gives us an independent check on our calibration:
comparisons of the PN magnitudes and line-of-sight velocities are
shown in Figures~\ref{f:n3379magcomp} and~\ref{f:n3379velcomp}.  The
magnitudes are consistent within their uncertainties, although there
is clearly crowding near the limiting magnitudes of either survey.
Considering that 50\% of our data was taken under non-photometric
conditions, the agreement is remarkable.  The additional tight
agreement between the two velocity measurements further convinces us
that we calibrated our data properly and adequately.

In our sample of PNe we have one candidate with an unusual velocity:
E27 at $291\mr{\ km\ s^{-1}}$.  It is unlikely to be associated with
NGC~3384, because of the latter's velocity field, which would favor
PNe with velocities over $720\mr{\ km\ s^{-1}}$.  Hence, it could
either be an interloper or an intragroup PN, but we have no way of
ascertaining either conjecture.  In the mass models we present for
NGC~3379, we exclude this object in our data analysis. 

The overall pattern of radial velocities (see Fig.~\ref{f:rvfields})
shows the signature of minor axis rotation.  Although there are not
enough data points for an informative independent estimate of the
velocity field based on the PNe alone, simple smoothing spline
estimates of the rotation and velocity dispersion (using code kindly
provided by D. Merritt) are consistent with the long slit results of
\citet{ss99}, yielding $V/\sigma\sim0.25$.  Hence we can apply the
Tracer Mass Estimator \citep[TME;][]{evanea03} to give a first
approximation of the total mass of NGC~3379.  Assuming a number
density $n\propto r^{-2.3}$ and a logarithmic potential, the TME
yields $(2.1\pm0.1) \times 10^{11}\mathrm{\ M_\sun}$ within 8~kpc,
where the uncertainties were derived assuming anisotropies
$\beta=\pm0.3$.  The implied $B$ band mass-to-light ratio is
$\Upsilon_B=9.9\pm0.5$.

\subsection{NGC~3384}

The SB0 galaxy NGC~3384 is the closest companion of NGC~3379 and is
one of the brightest members of the Leo (M96) group \citep{fs90}.
There is some evidence for interaction from a faint tidal arm
\citep{mal84} and the large HI ring that surrounds the galaxy pair
\citep{schn89}.  Three components contribute to the light distribution
of NGC~3384: a small bulge with a complex structure which includes the
bar within $\sim20\arcsec$, a lens that extends out to $160\arcsec$
and an outer exponential disk \citep{bea96}.  The galaxy was also
observed by \citet{cjf89}, who found a sample of 102 PNe (100 if we
identify two pairs of PNe separated by less than $0.5\arcsec$).  There
was, however, no spectroscopic follow up for their NGC~3384 sample.

NGC~3384 was observed with four pointings, giving an effective survey
area of $16.2\mr{\ arcmin^{-2}}$, which excludes the bright inner part
of the galaxy, ghosts, and the brightest reference stars.  Because the
RFP produces ghost images of bright objects the survey area has a
complicated topology; spectra will have as few as 19 data points in
some regions and as many as 65 data points in others.  We purposely
avoided the region between NGC~3384 and NGC~3379.  The galaxies have
systemic velocities that are close enough to make it impossible to
disentangle the correct host of each PN candidate in the intergalactic
area, even more considering the possible interaction between the two.
The NGC~3384 data have been discussed previously by \citet{tmw95}, but
was based on a different data reduction and candidate selection
process.  We present our sample of PNe in Table~\ref{t:3384-pne}; the
corresponding spectra are shown as
Figures~\ref{f:n1549spectra}.5--\ref{f:n1549spectra}.7 in the online
version of the Journal.

With our field of view we can, in principle, observe 82 PNe from the
\citet{cjf89} sample.  We recover 37 PNe and discovered 13 new PNe.
The comparison of the magnitudes of the PNe we have in common is shown
in figure~\ref{f:n3384magcomp}.  As before, the agreement is quite
good.  An interesting candidate is W05, the outlier at the bright end
of the plot.  Visual inspection shows possible structure around this
candidate.  \citet{cjf89} interpreted this as two close candidates
(their numbers 6 and 60), which seems unlikely since the two
candidates are close both in velocity space and on the sky.  Hence,
W05 could be an interloper, but we include it for completeness.

We can apply the TME to the NGC~3384 data, but we need to make some
additional assumptions, since Figure~\ref{f:rvfields}
shows the clear disk-like kinematics of the PN sample.  We will assume
that the S0 has a flat rotation curve at $120\mathrm{\ km\ s^{-1}}$
and that all PNe lie in the plane of the disk with a power law density
distribution.  The disk has a presumed inclination $i=62^\circ$; hence
we can split the line-of-sight velocities of the PNe into a component
due to the rotation and a random component.  We apply the TME to the
random component, adding $\langle v_\mathrm{rot} \rangle r/G$ for the
rotational motion to arrive at the total mass: $(5.1\pm0.3) \times
10^{10}\mathrm{\ M_\sun}$ within 9~kpc.  The implied $B$ band
$\Upsilon_B$ within this radius is $10.9\pm0.7$.

\section{Mass modeling}
\label{s:massmodels}

\subsection{Physical Framework}

Determining the potential of a spherical system, let alone an
axisymmetric or triaxial system, using a discrete kinematic tracer
population such as PNe would take on the order of $10^3$ radial
velocities \citep{ms93}.  With our samples ($N \lesssim 50$) clearly
falling short of this requirement, we shift the objective of our
modeling from finding the most likely mass model to assessing the
constraints the PNe velocities can place on the distribution of mass
at $r \gtrsim R_e$.  Only the two Leo galaxies, NGC~3379 and NGC~3384,
have PN sample sizes that are large enough to give potentially
interesting results.  In the following we consider spherical mass
models for the E1 galaxy NGC~3379; attempts to build axisymmetric mass
models for the S0 galaxy NGC~3384 were not successful and are not
presented here \citep{sl04}.

Our model space $M$ is predicated on the Jeans equation for a
spherical non-rotating system, assuming a distribution function of the
form $f(E,L^2)$:
\begin{equation}
   GM_<(r) = -r \sigma_r^2(r) \left( \frac{\mathrm{d} \ln
   \nu}{\mathrm{d} \ln r} + \frac{\mathrm{d} \ln
   \sigma_r^2}{\mathrm{d}\ln r} + 2\beta(r) \right),
   \label{e:sje}
\end{equation}
where $M_<(r)$ is the mass interior to radius $r$, $\nu(r)$ the
luminous density of the tracer population, $\sigma_r(r)$ the radial
velocity dispersion of said population, and $\beta(r) \equiv 1 -
\sigma_t^2(r) / \sigma_r^2(r)$ its velocity anisotropy.  The
gravitational potential of NGC~3379, an E1 galaxy, will be rounder
than its mass distribution \citep{bt87} and deviate from sphericity by
only a few per cent.  Because the PNe data sets are highly incomplete
within $1\ R_e$ we augment our kinematic data with the results of long
slit spectroscopy to constrain the models in the inner regions,
assuming both data sets represent the same tracer population.  In the
case of NGC~3379 we will use the results of \citet{ss99}.  The
kinematic data show an amount of rotation ($V/\sigma\sim0.25$) that is
slight enough to reasonably model the galaxy as a non-rotating system.

We derive the luminous density $\nu(r)$ from the surface brightness
$\mu(R)$ by applying an Abel inversion in the fashion of
\citet{gebea96}.  For the surface brightness $\mu_B(r)$ we combined
the groundbased data from \citet{pelea90} with the \textit{HST} data
(F555W filter) from \citet{gebea00} shifted to $B$ band by assuming a
uniform color \citep{gouea94}.  The combined surface brightness
profile only extends out to $154''$, and we performed a linear
extrapolation in $(\log r,\mu_B)$ for larger radii.

\subsection{Exploring the model space}

In looking for ways to analyze the range of plausible mass models for
NGC~3379, we insisted that our method be non-parametric and use the
data directly.  A non-parametric method, though generally
computationally intensive, will give a more conservative estimate of
the range of possible mass models and will come closer to the most
likely model than a parametric method.  Indeed, the form of a
parametric function can artificially constrain our search for
plausible mass models.  Furthermore, we want to be frugal with our
data and avoid unnecessary binning of the PNe velocities.  In other
words, instead of relying on the unstable process of estimating and
differentiating $\sigma_r^2(r)$ to infer $M_<(r)$ from
Equation~\ref{e:sje}, we posit $M_<(r)$ and measure its success in
matching the observed data.

The method developed by \citet[MB]{mb01}, which is briefly reiterated
below, meets both these requirements.  The MB analysis is Bayesian in
outlook and quantifies the plausibility of a mass model $M$ given the
data $D$:
\begin{equation}
   p(M|D) \propto p(D|M)p(M),
   \label{e:bt}
\end{equation}
where the ``prior'' probability $p(M)$ encodes our prejudices about
the most appropriate model $M$ absent any observations, the likelihood
(``chi-squared'') $p(D|M)$ quantifies the probability of observing the
data $D$ given a model $M$, and the ``posterior'' probability $p(M|D)$
quantifies the distribution of plausible models $M$ given our
prejudices and the data.  Credible regions---the Bayesian analog of
confidence regions---are naturally constructed from the posterior and
quantify the constraints the data place on the mass distribution.

We specify our models as follows.  Given a \emph{local} mass-to-light
profile $\Upsilon(r)$ to convert the luminous density $\nu(r)$ to a
mass density $\rho(r)$, and an anisotropy profile $\beta(r)$ to
determine $\sigma_t^2(r)$ from $\sigma_r^2(r)$, we can assay the
likelihood $p(D|M)$ of the kinematic data by integrating
equation~\ref{e:sje} to find the velocity dispersion along the line of
sight.  (In the case of the long slit data the projected velocity
dispersion is convolved with a Gaussian to mimic the effect of
seeing.)  The shape of $\Upsilon(r)$ is specified in terms of its
values $\Upsilon_i$ at a discrete set of sample points $r_i,
i=1,\ldots,N_\Upsilon$, with intermediate values obtained through
interpolation.  On the other hand, we will assume that the anisotropy
is constant with either $\beta=0$ or $\beta=0.3$, partly for ease of
computation, partly due to the work by \citet[fig.~4]{gerea01} who
find a remarkably constant anisotropy ($\beta=0.3$) in their dynamical
models of giant ellipticals.

Having specified the projection of our model into observable space, we
calculate the posterior probability $p(M|D)$ for a given trial
model $M = \{\Upsilon(r),\beta\}$ from the likelihood and the prior
probability.  We define the likelihood as follows
\begin{eqnarray}
   \ln p(D | \Upsilon(r),
   \beta) & = & -\frac{1}{2} \sum_{i} \left ( \frac{\sigma_i -
   \sigma_p(R_i)}{\Delta\sigma_i} \right)^2 \nonumber \\
    & & - \sum_{j}\left
   ( \frac{v_j^2}{2\sigma_p^2(R_j)}  -
   \ln \frac{\Delta v_j}{\sigma_p(R_j)} \right),
   \label{e:likelihood}
\end{eqnarray}
where $\sigma_i$ and $\Delta\sigma_i$ are the observed velocity
dispersions from the long slit spectra and their uncertainties, and
$v_j$ and $\Delta v_j$ are the radial velocities of the PNe and their
uncertainties.  The first term quantifies the goodness-of-fit for the
long slit data and the second term quantifies the likelihood of a
radial velocity $v_j$ given the velocity dispersion along the line of
sight.  The prior probability of a trial model is in essence a
smoothness constraint:
\begin{equation}
   \ln p(\Upsilon(r),\beta) = -\lambda \int \left[
   \frac{\dl{}^2(\ln\Upsilon)} {\dl(\ln r)^2} \right]^2 \dl (\ln r),
   \label{e:prior}
\end{equation}
where $\lambda$ specifies the amount of smoothness we prefer in our
trial $\Upsilon(r)$.  The prior gives a high probability to
mass-to-light profiles that are close to a power-law,
i.e. $\Upsilon(r) \sim r^\alpha$.  In our experience the results of
our calculation are not very sensitive to the value of $\lambda$: we
settled on $\lambda=0.1$.  Nor is the value critical, since it simply
encoded our prejudice as to what constitutes $\Upsilon(r)$ with too
much variation.  The next step is to investigate the posterior
$P(M|D)$.

At this point we deviate from the MB method.  Where they continue and
include an ansatz for the distribution function of the system to
ensure that the choices for $\Upsilon(r)$ correspond to a physical,
i.e. non-negative, distribution function, we decided to forego this
step in order to speed up the calculations.  Our results might not
always be physically plausible, but they are conservative, in the
sense that the physical results are at worst a subset of our results.

The high dimensionality of our model space ($N_\Upsilon=12$) precludes
sampling the posterior $p(M|D)$ using a simple grid.  Instead we
explore the model space using a Monte Carlo Markov Chain (MCMC)
method.  A MCMC is no more than a random walk which, given enough
steps, yields a set of trial models whose density distribution in the
model space is proportional to the posterior distribution.  The main
difficulty in using MCMCs is to ensure that the random walk has
wandered enough through model space for the density of trial models to
have converged to the posterior distribution.  To improve the
convergence of our MCMCs we use the slice sampler \citep{neal03} to
generate the random walk instead of the more conventional Metropolis
algorithm \citep{metrea53}.  Lacking a consensus among statisticians
about the correct way to establish the convergence of an MCMC we use
graphical checks of the random walks as well as the Gelman \& Rubin
$R$ statistic \citep[see, e.g.,][]{verdea03}.  For a more detailed
description of our implementation we refer the reader to \citet{sl04}.
 
\subsection{The results}

We performed two sets of MCMC calculations: one for isotropic
$(\beta=0)$ mass models and one for radially anisotropic $(\beta=0.3)$
mass models.  In both cases we sampled $\Upsilon(r)$ at $N_\Upsilon =
12$ logarithmically spaced radii.  We performed some exploratory
calculations with a range of values for $N_\Upsilon$.  Our chosen
value reflects the balance between an adequate resolution of
$\Upsilon(r)$ and a reasonable convergence speed of the MCMCs.  The
MCMC for the isotropic models required a total of $1.8\times10^5$
iterations to converge; the anisotropic models a total of
$4.6\times10^5$.  In both cases we discarded a total of $2\times 10^4$
iterations as a ``burn in'' period for the chains.  Once we have our
set of trial models it is straightforward to establish the, e.g., 95\%
credible region for a given $\Upsilon_i$ by determining the range in
$\Upsilon_i$ that encompasses 95\% of all the trial models.  Implicit
in this determination is the marginalization of the posterior
probability over all the other $\Upsilon_j, j\neq i$.  Likewise we
calculate the corresponding confidence regions for, say, $\sigma_p(R)$
and gauge how well our models can match the data.  The results of our
calculations are summarized in Figure~\ref{f:n3379_mcmc} with credible
regions of 99\%, 90\%, and 50\%.

To provide some context we obtained the best-fit models for three
parametric mass models: a Hernquist density profile $\propto r^{-4}$
\citep{her90}, an NFW profile $\propto r^{-3}$ \citep{nfw97}, and a
pseudo-isothermal profile $\propto r^{-2}$ \citep{bt87}, where in all
cases $r \gg r_s$ with $r_s$ the appropriate scale length; all three
models have a second parameter $\rho_s$ that scales the density.  The
best-fit models maximize the likelihood of the data---defined in
Equation~\ref{e:likelihood}---by varying $M_<(r|r_s,\rho_s)$, but
keeping $\nu(r)$ fixed, i.e. we do not assume a constant mass-to-light
ratio for these models.  They appear as the (red) curves in
Figure~\ref{f:n3379_mcmc}.

The bottom panels of Figure~\ref{f:n3379_mcmc} show that the
non-parametric description of $\Upsilon(r)$ has more leeway than the
parametric models in matching the actual data, especially around 1~kpc
where the two sides of the galaxy have significantly different
velocity dispersions and at the outer edges, where the PNe show a
steep drop in velocity dispersion.  We emphasize that our method does
not bin the PNe velocities; the figure shows binned PN data for
clarity.  The core radii of the parametric models are quite small, an
indicator of the failure of these mass models: the anisotropic
Hernquist model has the largest scalelength with 0.6~kpc, the
pseudo-isothermal model the smallest with 1.15~pc.  A
pseudo-isothermal profile, unsurprisingly, is not able to reproduce
the data, because of its nearly constant velocity dispersion; the
Hernquist and NFW profiles do better, but still show a clear bias,
justifying our choice for a non-parametric approach.

The difference between the two sets of models, at least in terms of
matching the data, is a widening of the credibility contours in the
case of the anisotropic models.  (The results from MB for NGC~3379 in
their Figure~3 show a similar widening.)  Whether or not this is a
significant difference can be answered by calculating the odds ratio
(see Equation~14 of MB) of the isotropic versus the anisotropic
hypothesis:
\begin{equation}
   O = \frac{p(\beta_1|D)}{p(\beta_2|D)}
     = \frac{p(\beta_1)}{p(\beta_2)} \frac{\int p(D|\Upsilon,\beta_1)
     p(\Upsilon) \mr{d} \Upsilon} {\int p(D|\Upsilon,\beta_2)
     p(\Upsilon) \mr{d} \Upsilon}. 
\end{equation}
Typically, an odds ratio $O\sim10$ is considered to be conclusive in
favoring one hypothesis over another.  Since we do not have any
\textit{a priori} preference for the value of $\beta$, the odds ratio
reduces to the ratio of the average values of the posterior
distributions.  In the case of the parametric models $O \sim 2$,
implying either hypothesis is equally likely.  From the MCMCs we find
$O = 4.6$ in favor of isotropy, which is a hint, but hardly
conclusive.  As MB noted, this is not a surprising result: barring
information about the higher order velocity moments $\Upsilon(r)$ has
enough freedom to match to data, given some value (or even profile)
for $\beta$.  The little information of these higher order moments
encoded in the PN velocities is clearly not enough to be a useful
discriminant. 

The circular velocity in both sets of models is flat between 1 and
4~kpc at roughly $250\ \rm{km\ s^{-1}}$, but drops of rapidly outside
the latter radius, matching the results of
\citet[Figure~18]{kronea00}.  The agreement with the analysis of MB is
less convincing at large radii, a consequence of their use of the
smaller \citet{cjd93} PN sample.  The largest difference between the
two sets is within 0.6~kpc, where the anisotropic models need
significantly less mass than the isotropic models to produce the same
velocity dispersion profile.  This is simply a reflection of
degeneracy between mass and velocity anisotropy, which can only be
broken by the inclusion of higher order moments velocity data
\citep[see][]{mer93, ger93, vdmf93}.

The top panels of Figure~\ref{f:n3379_mcmc} show a constant
$\Upsilon(r)$ between 0.1 and 2~kpc, before showing a small hump
followed by a steep decrease at larger radii.  The range in
$\Upsilon(r)$ is consistent with the range found by \citet{gerea01}
for NGC~3379 from stellar population synthesis.  The bump is related
to the diverging velocity dispersion data points outside 1~kpc and
creates a small plateau in the model velocity dispersion.  The flaring
of the $\Upsilon(r)$ credibility contours shows that the constraints
on the mass-to-light profile from the PNe are weaker than those from
the long slit spectra, but are nonetheless informative within 10~kpc.
Compare the parametric models with their associated mass-to-light
profiles: they agree fairly well with $\Upsilon(r)$ within 1~kpc (long
slit spectra), but diverge strongly in the region dominated by the PN
data.  Despite this fact, the parametric models only allow us to
conclude that the mass density should fall off steeper than $\propto
r^{-2}$, but cannot discriminate between a Hernquist or an NFW
profile.  The non-parametric estimate $\Upsilon(r)$, however, not only
matches the data better, but also gives us hints of a breakdown of our
assumptions (a plateau instead of a flaring of confidence contours).
Most remarkable, however, remains the relative flatness of
$\Upsilon(r)$.  The implied total mass within 8~kpc is $7.5\times
10^{11}\mathrm{\ M_\sun}$, a factor of a few smaller than the TME
estimate in section~\ref{s:samp3379}.  The total mass-to-light ratio
implied is $\Upsilon_B\sim7$, consistent with the stellar population
models of \citet{gerea01}.

\subsection{Discussion}

The conclusion that our data show little evidence for DM in the inner
8~kpc of NGC~3379 is hardly surprising in the light of the work done
by \citet{cjd93}, \citet{kronea00}, and---more
recently---\citet{romea03}.  In particular, a cursory inspection of
the PN data in Figure~3 of \citet{cjd93} shows a striking resemblance
to the similar data in our Figure~\ref{f:n3379_mcmc}: a steep drop off
in velocity dispersion between $1R_e$ and $3R_e$.  Using a parametric
model that guaranteed a positive distribution function they found no
significant signature for the presence of DM within $3.5R_e$ with $M =
1 \times 10^{11}\mathrm{\ M_\sun}$ and $\Upsilon_B\sim7$ consistent
with our own findings.  The investigations by \citet{kronea00}, based
on absorption line spectra out to $\sim2R_e$, arrive at the same but
more general conclusion: ellipticals likely have nearly maximal
mass-to-light ratios.  Specifically, they find $\Upsilon_B=4.5$.

On the basis of orbit superposition modeling \citet{romea03}
determined a mass-to-light ratio for NGC~3379 in the $B$ band of
$7.1\pm0.6$.  Their sample included about $\sim100$ PNe and also
showed the radial decline in velocity dispersion.  Indeed, they found
the same decline in three more galaxies and concluded that elliptical
galaxies contain little DM when compared to other galaxies.
Considering the range in modeling procedures employed to study the
dynamics of NGC~3379 and the consistency of the results, it is fair to
say that it shows no evidence for dark matter within the inner few
effective radii.  At the same time, analysis of the HI~ring around
NGC~3379 and NGC~3384 implies an enclosed mass of $\sim
6\times10^{11}\mathrm{\ M_\sun}$ \citep{schnea89}.  The discrepancy
between the total dynamical mass-to-light ratio $\Upsilon_B=27$ within
the 110~kpc ring \citep{schn89} and within a 10~kpc radius of NGC~3379
suggests that most of the dark matter is at large radii.

Our models might not capture the complexity of NGC~3379 acceptably.
The galaxy is not perfectly spherical and might even be an S0 seen
face on \citep{cea91}, but as \citet{romea03} have observed, either
possibility is unlikely to have a large effect on our final answer.
The fact, noted by \citet{ss99}, that the long slit data show a bump
in the $h_3$ moment at $\sim 15''$ is interesting in this regard as an
indication of a more complex description for NGC~3379 than
incorporated in our model.  The twists in the photometric surface
brightness and the kinematic velocity field (both roughly 5 degrees,
but in opposite directions) already hint at such a departure from our
assumptions.  \citep{ss99} suggest that NGC~3379 might be triaxial.
The system might also not have adequately relaxed.  Indeed, the
asymmetry of the velocity dispersion profile beyond $20''$ seems to
corroborate the latter possibility, despite the lack of photometric
evidence.  

A more worrisome development, from a recent paper \citet{sgm05}, is
the discovery of two populations of PNe in NGC~4697, one younger and
inherently brighter than the other, creating a bias in the
measurements of the PN kinematics.  Given its somewhat odd velocity
field, it is entirely possible that NGC~3379 contains a bimodal
population of PNe.  A sample size of 50~PNe, however, is clearly
unable to address these questions usefully.  A larger sample of
ellipticals combined with deeper observations are needed to
investigate the relevant statistics.

\section{Conclusions}
\label{s:wrapup}

We have reported on a search for PNe around four early-type galaxies
using the RFP.  We obtained reasonably sized samples in the case of
the two Leo galaxies, NGC~3379 and NGC~3384, with adequate photometry
and well determined radial velocities.  The main limiting factor in
our observations was the seeing; our data are read noise limited and
better seeing would have improved the limiting magnitude of our
survey.  In our analysis of the RFP data cubes two realizations were
essential in optimizing the size of the extracted PN samples.  The
power of our detections comes from using the data cube as a whole, as
opposed to looking for point sources in each monochromatic RFP image.
Secondly, applying the proper statistic is essential, along with a
proper characterization of the reference distribution.  A simple
estimator yields mass-to-light ratios $\Upsilon_B \sim 10$ for the Leo
galaxies, although a more sophisticated analysis for NGC~3379 gives an
estimate that is a factor two lower, making it consistent with stellar
population mass-to-light ratios.  Although our models are relatively
simple, they do produce conservative estimates of the mass-to-light
ratios, and hence we do not find evidence for a dominant DM component
inside a few $R_e$, confirming the recent work by \citet{romea03}.

In order to address questions about the existence of multimodal PN
populations or the properties of PN kinematics as a function of host
galaxy luminosity, we clearly need a larger and deeper set of surveyed
galaxies.  The RFP has been decommissioned, but successor instruments
will be coming online in the very near future.  The Prime Focus
Imaging Spectrograph on the Southern African Large Telescope (11~m)
will have Fabry-Perot image spectroscopy as one of its modes and will
be ideally suited to extend current surveys of extragalactic PNe.  It
will be both competitive, considering depth and field of view, and
complementary, being on the Southern hemisphere, to the Planetary
Nebulae Spectrograph \citep{dea02}.

\acknowledgements

Benoit Tremblay collaborated in some of the observations of this
study.  The staff at CTIO provided their usual excellent support for
these observations.  AS would like to acknowledge helpful
conversations and email exchanges with E.~Barnes, D.~Chakrabarty,
R.~M\'endez, R.~Ciardullo, and J.Magorrian.  This work was supported
in part by the National Science Foundation through grants AST9731052
and AST0098650.

{\it Facilities:} \facility{Blanco (RFP), \facility{CTIO:1.5m}}



\begin{deluxetable}{lll}

\tablewidth{0pt} 
\tablecaption{The instrumental and detector setup. \label{t:setup}}
\tablehead{
\colhead{} & \colhead{1994 April 7--10} & \colhead{1995 February 2--6} }

\startdata
CCD   & Tek 1k (CTIO \#1, VEB) & Tek 1k (CTIO \#2, Arcon) \\
Gain  & 1.73 $e^-\ \mr{ADU^{-1}}$    & 1.25 $e^-\ \mr{ADU^{-1}}$ \\
Read noise  & 3.22 $e^-$          & 4.33 $e^-$ \\
Image scale & 0\farcs70 $\mr{pix^{-1}}$ & 0\farcs35 $\mr{pix^{-1}}$ \\
Voigt FWHM                     & 2.0\phn~\AA   & 2.1\phn~\AA \\
---Gaussian width $\Delta\lambda_G$   & 0.41~\AA  & 0.22~\AA \\
---Lorentzian width $\Delta\lambda_L$ & 0.89~\AA  & 1.0\phn~\AA \\
Free spectral range $\Delta\lambda_\mathrm{FSR}$ & 48~\AA & 48~\AA \\
Filters (at CTIO) & 5007-44, 5037-44 & 5007-44, 5037-44  \\
Switching wavelength $\lambda_\mr{s}$ & 5019~\AA & 5022~\AA \\
\enddata

\end{deluxetable}



\begin{deluxetable}{lllcccccll}

\setlength{\tabcolsep}{0.03in}
\tablewidth{0pt} 
\tablecaption{The observed fields. \label{t:fields}}
\tablecolumns{8}
\tablehead{
\colhead{field} & \colhead{$\alpha_c$} & \colhead{$\delta_c$} &
\colhead{offset} & \colhead{$N_\mathrm{obs}$}\tablenotemark{a} & 
\colhead{$\Delta\lambda$} & \colhead{$<\!\delta\lambda\!>$}\tablenotemark{b} &
\colhead{$N_\mathrm{stars}$} & \colhead{seeing}\tablenotemark{c} &
\colhead{$m_\mr{lim}$} \\
\colhead{} & \colhead{(hms)} & \colhead{(\degr\,\arcmin\,\arcsec)} &
\colhead{(\arcsec)} & \colhead{} &
\colhead{(\AA)} & \colhead{(\AA)} & \colhead{} & \colhead{(\arcsec)} &
\colhead{} \\
}

\startdata
NGC~1549  & 04 15 45&$-55$ 35 27&   6& 28 (5)& 5016--5043& 1.0& 3& 1.5*& 26.3\\
NGC~3379 E& 10 47 54& +12 35 27 &  69& 29 (2)& 5011--5039& 1.0& 3& 1.6*& 26.1\\
NGC~3379 W& 10 47 45& +12 34 25 &  77& 33 (2)& 5011--5040& 1.0& 3& 1.5*& 26.1\\
NGC~3384 C& 10 48 17& +12 37 42 &   4& 27 (2)& 5012--5035& 0.6& 5& 1.8 & 26.2\\
NGC~3384 N& 10 48 17& +12 38 14 &  31& 19 (1)& 5012--5035& 1.2& 5& 1.7 & 26.1\\
NGC~3384 W& 10 48 12& +12 37 47 &  75& 19 (1)& 5012--5035& 1.2& 4& 2.0 & 26.2\\
NGC~3384 E& 10 48 26& +12 39 34 & 171& 19 (1)& 5010--5028& 1.0& 6& 1.6 & 25.9\\
NGC~4636  & 12 42 50& +02 41 15 &   4& 34 (4)& 5013--5034& 0.6& 3& 2.2*& 26.7\\
\enddata

\tablecomments{NGC~1549, NGC~3379 and NGC~3384~E were observed during
the 1995 run, the other fields during the 1994 run.  For the 1994
(1995) run we took dome flats at 21 (28) RFP settings, spaced
1.25~\AA\ (1.0~\AA) apart, in sets of three images with an exposure
time of 120~s (300~s) which were averaged to produce the final
flatfield image.}

\tablenotetext{a}{The number of images in the RFP stack with the
  number of observing nights in parentheses.}

\tablenotetext{b}{The average wavelength separation of images in the
  RFP stack.}

\tablenotetext{c}{An asterisk indicates that some of the data were
taken under non-photometric conditions.}

\end{deluxetable}



\begin{deluxetable}{lcccc}

\tablewidth{0pt} 
\tablecaption{Basic properties of the sample galaxies. \label{t:galsamp}}
\tablehead{
\colhead{} & \colhead{NGC 1549} & \colhead{NGC 3379} & \colhead{NGC 3384} &
\colhead{NGC 4636}
}

\startdata
Right ascension (2000.0; hms)\tablenotemark{a}               
& 04 15 44.00    & 10 47 49.60     & 10 48 16.90     & 12 42 49.87 \\
Declination (2000.0; $\degr\;\arcmin\;\arcsec$)\tablenotemark{a}  
&-55 35 30.0\phd & 12 34 53.9\phd  &  12 37 45.5\phd &  02 41 16.0\phd \\
Type\tablenotemark{b}            & E/S$0_1$ & E1 & SB(s)0- & E/S$0_1$ \\
Radial velocity $(\mathrm{km\ s^{-1}})$\tablenotemark{a}
& 1220 & 911 & 704 & 938 \\
Distance (Mpc)\tablenotemark{c} & 19.7 & 11.1 & 11.4 & 15.0\\
Total $B$ magnitude\tablenotemark{a} &  10.72 & 10.24 & 10.85 & 10.34
\\
$M_{B_T}$\tablenotemark{a} & -20.75 & -19.99 & -19.43 & -20.54 \\ 
$R_e$ ($\arcsec$)\tablenotemark{b}  & 91 & 35 & 50 & 177 \\
\enddata

\tablenotetext{a}{From the NASA/IPAC Extragalactic DataBase (NED).}
\tablenotetext{b}{Galaxy classifications from RC3 \citep{rc3}.}
\tablenotetext{c}{Using the mean distance modulus from \citet{f00},
apart from NGC~1549 \citep{tea01}.}
\end{deluxetable}



\begin{deluxetable}{lccrrrrrrrr}


\setlength{\tabcolsep}{0.03in}
\tabletypesize{\small}
\tablewidth{0pt} 
\tablecaption{NGC 1549 and NGC 4636 Planetary Nebulae \label{t:1549-pne}}
\tablecolumns{11}

\tablehead{
   \colhead{ID} & \colhead{$\alpha$} & \colhead{$\delta$} &
   \colhead{$\lambda_c$} & \colhead{$v_\mr{helio}$} &
   \colhead{$\sigma_v$} & \colhead{$F_\mr{pn}$} &
   \colhead{$m_{5007}$} & \colhead{$\sigma_m$} & \colhead{$S$} &
   \colhead{$Q$} \\
   \colhead{(1)} &\colhead{(2)} & \colhead{(3)} & \colhead{(4)} &  
   \colhead{(5)} &\colhead{(6)} & \colhead{(7)} & \colhead{(8)} &  
   \colhead{(9)} &\colhead{(10)} & \colhead{(11)} 
}

\startdata
\cutinhead{NGC 1549}
N1-1& 4:15:35.96& -55:35:18.3& 5026.2& 1147& 14& 1.04& 26.21& 0.23&  23.1&  5.0\\
N1-2& 4:15:42.65& -55:34:43.5& 5032.9& 1547& 17& 0.84& 26.45& 0.23&  22.8&  5.7\\
N1-3& 4:15:44.86& -55:36:10.7& 5033.8& 1603& 18& 0.98& 26.28& 0.25&  19.6&  5.6\\
N1-4& 4:15:48.31& -55:36:01.9& 5024.7& 1060& 14& 1.20& 26.07& 0.22&  26.7&  5.8\\
N1-5& 4:15:51.14& -55:34:34.8& 5025.1& 1080& 15& 0.93& 26.34& 0.20&  31.9&  6.4\\
N1-6& 4:15:51.75& -55:35:23.0& 5027.6& 1234& 12& 1.23& 26.03& 0.19&  34.1&  7.5\\
\cutinhead{NGC 4636}
N4-1& 12:42:51.53&  2:40:49.9& 5019.3&  739& 19& 0.77& 26.54& 0.26& 17.4&  4.8\\
N4-2& 12:42:52.09&  2:40:24.9& 5025.1& 1086& 21& 0.75& 26.58& 0.23& 23.3&  7.1\\
\enddata			       

\tablecomments{Col. (1) N1 and N4 denote PNe found in NGC~1549 and
  NGC~4636, respectively; col. (2)--(3) right ascension (hms) and
  declination ($\degr\arcmin\arcsec$), where the uncertainty in each
  coordinate is $0\farcs7$ (includes random and systematic
  uncertainty); col. (4) peak wavelength (\AA); col. (5) heliocentric
  line-of-sight velocity ($\mr{km\ s^{-1}}$); col. (6) uncertainty in
  $v_\mr{helio}$; col. (7) emission line flux ($10^{-16}\mr{\ erg\
  cm^{-2}\ s^{-1}}$); col. (8) emission line magnitude $m_{5007}\equiv
  -2.5\log(F)+13.74$ (no extinction correction was made); col. (9)
  uncertainty in $m_{5007}$; col. (10) test statistic $S_\mr{LR}$;
  col. (11) signal-to-noise ratio. }

\end{deluxetable}



\begin{deluxetable}{lccrrrrrrrrrrr}


\setlength{\tabcolsep}{0.03in}
\tabletypesize{\small}
\tablewidth{0pt} 
\tablecaption{NGC 3379 Planetary Nebulae \label{t:3379-pne}}
\tablecolumns{11}

\tablehead{
   \colhead{ID} & \colhead{$\alpha$} & \colhead{$\delta$} &
   \colhead{$\lambda_c$} & \colhead{$v_\mr{helio}$} &
   \colhead{$\sigma_v$} & \colhead{$F_\mr{pn}$} &
   \colhead{$m_{5007}$} & \colhead{$\sigma_m$} & \colhead{$S$} &
   \colhead{$Q$} & \colhead{$\mr{ID_{cjf}}$} &
   \colhead{$m_\mr{cjf}$} & \colhead{$v_\mr{cjf}$} \\
   \colhead{(1)} &\colhead{(2)} & \colhead{(3)} & \colhead{(4)} &  
   \colhead{(5)} &\colhead{(6)} & \colhead{(7)} & \colhead{(8)} &  
   \colhead{(9)} &\colhead{(10)} & \colhead{(11)} & \colhead{(12)} &  
   \colhead{(13)} &\colhead{(14)}
}

\startdata
E01& 10:47:48.73& 12:35:24.8& 5024.8& 1083&  13& 1.68& 25.69& 0.21&  27.5&  6.0& ..&  ... &  ...\\
E02& 10:47:49.17& 12:35:42.4& 5017.6&  653&  17& 1.61& 25.74& 0.23&  26.3&  5.5& ..&  ... &  ...\\
E03& 10:47:49.47& 12:35:35.3& 5023.6& 1009&  17& 1.52& 25.80& 0.19&  32.6&  7.0&  9& 25.66&  ...\\
E04& 10:47:49.66& 12:36:02.5& 5021.0&  853&  18& 1.11& 26.14& 0.23&  23.4&  5.8& ..&  ... &  ...\\
E05& 10:47:49.94& 12:35:40.2& 5018.3&  694&  13& 1.97& 25.52& 0.22&  26.3&  6.1& 20& 25.84&  ...\\
E06& 10:47:50.30& 12:35:13.0& 5025.0& 1098&  12& 2.03& 25.49& 0.18&  37.1&  6.7& ..&  ... &  ...\\
E07& 10:47:51.06& 12:35:01.5& 5023.0&  974&  17& 1.78& 25.64& 0.22&  25.4&  6.9& 12& 25.76&  ...\\
E08& 10:47:51.15& 12:35:15.4& 5018.8&  723&  19& 1.85& 25.59& 0.26&  18.8&  4.8& 36& 26.10&  ...\\
E09& 10:47:51.26& 12:34:51.2& 5025.8& 1141&  17& 1.73& 25.66& 0.21&  26.4&  6.3& ..&  ... &  ...\\
E10& 10:47:51.40& 12:35:31.5& 5023.8& 1022&  13& 2.26& 25.37& 0.15&  54.7&  9.0&  2& 25.33& 1061\\
E11& 10:47:51.59& 12:35:43.3& 5025.4& 1116&  13& 1.73& 25.66& 0.15&  53.4&  8.2&  6& 25.53&  ...\\
E12& 10:47:51.68& 12:34:47.3& 5023.6& 1013&  10& 2.44& 25.29& 0.17&  39.3&  6.9& 23& 25.92&  ...\\
E13& 10:47:52.03& 12:35:11.0& 5018.0&  675&  20& 1.76& 25.64& 0.22&  25.3&  4.7& ..&  ... &  ...\\
E14& 10:47:52.11& 12:34:43.5& 5022.4&  942&  13& 2.19& 25.41& 0.15&  50.3&  9.1&  1& 25.28&  936\\
E15& 10:47:52.28& 12:35:40.1& 5023.3&  991&  13& 1.81& 25.62& 0.17&  45.1&  8.0& 14& 25.77&  ...\\
E16& 10:47:52.33& 12:36:05.7& 5020.8&  844&  11& 1.75& 25.65& 0.18&  35.6&  6.7& 16& 25.78&  832\\
E17& 10:47:53.34& 12:36:17.2& 5021.4&  881&  14& 1.25& 26.02& 0.22&  23.5&  6.0& ..&  ... &  ...\\
E18& 10:47:54.83& 12:36:24.8& 5020.8&  843&  13& 1.47& 25.84& 0.20&  31.0&  5.6& 30& 26.00&  ...\\
E19& 10:47:55.12& 12:36:17.1& 5021.3&  873&  13& 1.43& 25.87& 0.20&  28.8&  6.7& 60& 26.37&  ...\\
E20& 10:47:55.43& 12:34:51.5& 5023.5& 1007&  17& 1.28& 25.99& 0.22&  24.2&  6.0& 38& 26.14& 1021\\
E21& 10:47:55.50& 12:34:46.0& 5020.9&  850&  13& 1.57& 25.77& 0.21&  28.3&  6.6& ..&  ... &  ...\\
E22& 10:47:55.76& 12:35:40.5& 5020.2&  806&  19& 1.43& 25.87& 0.27&  17.0&  4.9& ..&  ... &  ...\\
E23& 10:47:56.07& 12:35:42.0& 5015.2&  510&  17& 1.46& 25.85& 0.22&  23.8&  5.4& ..&  ... &  ...\\
E24& 10:47:56.86& 12:36:14.5& 5019.2&  748&  21& 1.16& 26.10& 0.27&  17.0&  4.3& 28& 25.98&  ...\\
E25& 10:47:56.90& 12:34:20.7& 5015.4&  521&  17& 1.30& 25.97& 0.27&  15.8&  4.6& ..&  ... &  ...\\
E26& 10:47:56.92& 12:35:31.2& 5021.9&  910&  14& 1.36& 25.93& 0.20&  28.5&  6.2& 39& 26.15&  ...\\
E27& 10:47:58.10& 12:34:42.2& 5011.6&  291&  23& 1.71& 25.68& 0.30&  22.1&  5.4& ..&  ... &  ...\\
E28& 10:47:59.21& 12:35:10.6& 5018.7&  719&  13& 1.79& 25.63& 0.20&  29.8&  6.8& 10& 25.68&  ...\\
W01& 10:47:41.41& 12:35:12.0& 5020.0&  797&  17& 1.09& 26.17& 0.22&  23.9&  3.6& ..&  ... &  ...\\
W02& 10:47:41.67& 12:34:05.5& 5018.7&  720&  17& 1.15& 26.11& 0.23&  22.7&  5.5& ..&  ... &  ...\\
W03& 10:47:41.71& 12:34:32.2& 5023.4& 1004&  15& 1.32& 25.96& 0.20&  28.4&  6.9& 48& 26.26&  ...\\
W04& 10:47:42.55& 12:35:31.7& 5015.3&  517&  17& 1.12& 26.14& 0.25&  18.6&  4.3& ..&  ... &  ...\\
W05& 10:47:42.56& 12:35:08.9& 5023.8& 1025&  13& 1.39& 25.91& 0.15&  52.0&  8.9& 41& 26.18&  ...\\
W06& 10:47:42.92& 12:35:09.6& 5022.0&  919&  15& 1.25& 26.02& 0.21&  28.1&  6.8& 62& 26.40&  ...\\
W07& 10:47:44.73& 12:35:07.6& 5024.0& 1039&  12& 1.17& 26.09& 0.19&  34.7&  7.4& ..&  ... &  ...\\
W08& 10:47:45.21& 12:34:59.7& 5023.1&  982&  12& 1.48& 25.83& 0.17&  39.5&  7.2& 29& 25.99&  977\\
W09& 10:47:45.26& 12:35:17.3& 5018.7&  720&  11& 1.50& 25.82& 0.16&  48.2&  8.6& 19& 25.83&  716\\
W10& 10:47:45.39& 12:33:37.9& 5018.2&  691&  18& 1.24& 26.03& 0.23&  25.0&  4.7& 50& 26.28&  726\\
W11& 10:47:45.67& 12:34:13.4& 5017.4&  640&  18& 1.36& 25.93& 0.23&  25.2&  4.9& 63& 26.40&  ...\\
W12& 10:47:45.83& 12:35:06.7& 5025.7& 1138&  15& 1.06& 26.19& 0.21&  27.6&  7.3& ..&  ... &  ...\\
W13& 10:47:46.23& 12:33:37.5& 5026.9& 1212&  16& 1.02& 26.24& 0.24&  20.4&  5.5& 40& 26.15&  ...\\
W14& 10:47:46.24& 12:34:10.5& 5024.7& 1078&  15& 1.08& 26.18& 0.22&  25.0&  5.7& 59& 26.37&  ...\\
W15& 10:47:46.72& 12:33:18.7& 5015.6&  531&  20& 1.16& 26.10& 0.26&  17.9&  4.3& ..&  ... &  ...\\
W16& 10:47:46.94& 12:34:06.0& 5024.2& 1050&  16& 1.01& 26.25& 0.22&  24.0&  4.4& ..&  ... &  ...\\
W17& 10:47:47.25& 12:35:01.3& 5028.6& 1310&  11& 1.53& 25.80& 0.18&  36.2&  6.9& 11& 25.75&  ...\\
W18& 10:47:47.29& 12:34:47.1& 5026.2& 1168&  12& 1.48& 25.83& 0.18&  37.6&  7.4& ..&  ... &  ...\\
W19& 10:47:48.10& 12:35:00.1& 5020.0&  797&  13& 1.54& 25.79& 0.23&  22.2&  5.9& ..&  ... &  ...\\
W20& 10:47:48.34& 12:34:38.0& 5024.3& 1053&  13& 1.91& 25.56& 0.15&  54.0&  9.8&  7& 25.63& 1060\\
W21& 10:47:48.58& 12:33:32.9& 5019.3&  757&  18& 1.37& 25.92& 0.21&  27.0&  5.5& 49& 26.28&  ...\\
W22& 10:47:48.65& 12:35:03.5& 5022.3&  932&  10& 2.40& 25.31& 0.14&  61.5&  9.4& ..&  ... &  ...\\
W23& 10:47:48.85& 12:34:41.6& 5020.2&  807&  17& 1.58& 25.76& 0.25&  21.4&  4.5& ..&  ... &  ...\\
W24& 10:47:48.94& 12:33:40.3& 5023.4& 1001&  14& 1.22& 26.04& 0.19&  33.4&  6.3& 27& 25.96&  985\\
W25& 10:47:48.97& 12:34:27.6& 5019.4&  758&  15& 1.67& 25.70& 0.18&  39.0&  7.3& 42& 26.20&  ...\\
W26& 10:47:49.97& 12:34:38.8& 5021.0&  858&  16& 1.95& 25.54& 0.21&  30.6&  5.9& ..&  ... &  ...\\
\enddata			       

\tablecomments{Col. (1) E and W denote PNe found in the East and West
fields, respectively; col. (2)--(3) right ascension (hms) and
declination ($\degr\arcmin\arcsec$), where the uncertainty in each
coordinate is $0\farcs4$ (includes random and systematic uncertainty);
col. (4) peak wavelength (\AA); col. (5) heliocentric line-of-sight
velocity ($\mr{km\ s^{-1}}$); (6) uncertainty in $v_\mr{helio}$;
col. (7) emission line flux ($10^{-16}\mr{\ erg\ cm^{-2}\ s^{-1}}$);
col. (8) emission line magnitude $m_{5007}\equiv -2.5\log(F)+13.74$
(no extinction correction was made); col. (9) uncertainty in
$m_{5007}$; col. (10) test statistic $S_\mr{LR}$; col. (11)
signal-to-noise ratio; col. (12)--(13) corresponding
ID and $m_{5007}$ from \citet{cjf89}; col. (14) corresponding
line-of-sight velocity from \citet{cjd93}. }

\end{deluxetable}



\begin{deluxetable}{lccrrrrrrrrrrr}


\setlength{\tabcolsep}{0.03in}
\tabletypesize{\small}
\tablewidth{0pt} 
\tablecaption{NGC 3384 Planetary Nebulae \label{t:3384-pne}}
\tablecolumns{11}

\tablehead{
   \colhead{ID} & \colhead{$\alpha$} & \colhead{$\delta$} &
   \colhead{$\lambda_c$} & \colhead{$v_\mr{helio}$} &
   \colhead{$\sigma_v$} & \colhead{$F_\mr{pn}$} &
   \colhead{$m_{5007}$} & \colhead{$\sigma_m$} & \colhead{$S$}\tablenotemark{a} &
   \colhead{$Q$} & \colhead{$\mr{ID_{cjf}}$} &
   \colhead{$m_\mr{cjf}$} \\
   \colhead{(1)} &\colhead{(2)} & \colhead{(3)} & \colhead{(4)} &  
   \colhead{(5)} &\colhead{(6)} & \colhead{(7)} & \colhead{(8)} &  
   \colhead{(9)} &\colhead{(10)} & \colhead{(11)} & \colhead{(12)} &  
   \colhead{(13)} 
}

\startdata
E01& 10:48:21.63& 12:38:43.9& 5015.2& 497& 14& 1.32& 25.96& 0.21&  27.5 (19)&  5.8& 49& 26.38\\
E02& 10:48:24.42& 12:38:43.6& 5016.5& 576& 11& 1.55& 25.78& 0.18&  38.0 (19)&  6.8&  1& 25.59\\
E03& 10:48:25.18& 12:39:36.2& 5016.3& 564& 13& 1.07& 26.19& 0.23&  23.4 (19)&  6.4& 11& 25.93\\
W01& 10:48:09.00& 12:37:44.6& 5020.8& 812& 13& 1.10& 26.16& 0.24&  21.4 (19)&  6.7& 21& 26.05\\
W02& 10:48:09.63& 12:37:36.7& 5021.2& 838&  8& 1.79& 25.63& 0.16&  48.9 (18)&  9.8& 10& 25.90\\
W03& 10:48:10.43& 12:36:54.5& 5021.3& 844& 12& 1.31& 25.97& 0.21&  26.8 (19)&  8.2& 56& 26.43\\
W04& 10:48:11.56& 12:37:32.9& 5020.6& 800& 10& 1.35& 25.93& 0.14&  61.5 (45)& 12.4&  7& 25.81\\
W05& 10:48:11.85& 12:37:21.2& 5021.5& 857&  5& 2.58& 25.23& 0.08& 172.6 (44)& 17.1&  6& 25.73\\
W06& 10:48:11.97& 12:36:27.4& 5022.0& 886&  8& 1.90& 25.56& 0.16&  48.6 (18)&  9.5&  8& 25.85\\
W07& 10:48:12.29& 12:36:53.1& 5021.2& 838& 17& 0.90& 26.37& 0.20&  28.8 (35)&  8.4& 68& 26.58\\
W08& 10:48:12.80& 12:36:53.8& 5022.2& 896&  8& 1.92& 25.55& 0.11& 107.0 (42)& 16.6&  2& 25.63\\
W09& 10:48:13.25& 12:38:02.3& 5022.6& 920& 12& 0.80& 26.51& 0.17&  39.4 (61)&  9.6& 46& 26.36\\
W10& 10:48:13.28& 12:37:52.1& 5020.6& 800&  9& 0.95& 26.32& 0.16&  45.8 (62)&  9.6& 43& 26.34\\
W11& 10:48:13.41& 12:38:06.9& 5018.9& 698& 13& 0.92& 26.36& 0.18&  37.2 (61)&  8.7& 12& 25.94\\
W12& 10:48:13.49& 12:37:10.5& 5021.0& 826& 11& 1.29& 25.99& 0.15&  59.9 (51)& 10.9& 28& 26.11\\
W13& 10:48:13.51& 12:37:02.8& 5022.6& 923& 14& 0.86& 26.42& 0.23&  24.3 (44)&  7.2& 14& 25.96\\
W14& 10:48:13.64& 12:36:46.1& 5021.9& 883&  7& 1.88& 25.57& 0.10& 112.8 (42)& 15.0&  9& 25.86\\
W15& 10:48:14.17& 12:36:39.0& 5019.8& 754& 17& 0.86& 26.43& 0.21&  26.4 (45)&  6.9& ..&  ... \\
W16& 10:48:14.25& 12:37:04.6& 5020.6& 800& 11& 1.01& 26.25& 0.17&  39.5 (63)&  8.9& 59& 26.45\\
W17& 10:48:14.37& 12:36:47.7& 5021.3& 842& 11& 1.01& 26.25& 0.18&  34.6 (44)&  8.2& 37& 26.30\\
W18& 10:48:14.38& 12:37:23.9& 5021.0& 826& 11& 1.07& 26.19& 0.16&  45.3 (64)& 10.5& ..&  ... \\
W19& 10:48:14.60& 12:36:59.9& 5021.2& 839& 11& 1.23& 26.03& 0.17&  39.4 (47)&  8.3& ..&  ... \\
W20& 10:48:14.87& 12:37:25.8& 5021.4& 847&  9& 1.65& 25.71& 0.11&  93.2 (62)& 13.5&  3& 25.66\\
W21& 10:48:15.12& 12:38:08.3& 5019.5& 735& 11& 0.99& 26.28& 0.19&  33.0 (62)&  8.9& 94& 26.96\\
W22& 10:48:16.00& 12:37:03.3& 5020.3& 781& 14& 0.78& 26.54& 0.22&  22.9 (64)&  7.1& ..&  ... \\
W23& 10:48:16.03& 12:38:02.9& 5020.9& 821& 11& 1.22& 26.04& 0.18&  36.9 (61)&  8.4& 24& 26.07\\
W24& 10:48:16.12& 12:37:17.0& 5023.2& 956&  9& 1.16& 26.10& 0.16&  46.3 (63)& 10.9& 34& 26.20\\
W25& 10:48:16.17& 12:37:09.2& 5021.0& 824& 14& 0.94& 26.32& 0.21&  27.5 (61)&  9.1& ..&  ... \\
W26& 10:48:16.55& 12:38:09.9& 5019.8& 752& 17& 0.93& 26.34& 0.20&  28.4 (59)&  8.1& 33& 26.18\\
W27& 10:48:16.81& 12:37:20.1& 5023.1& 951& 11& 1.42& 25.88& 0.16&  45.9 (59)& 10.4& 18& 25.99\\
W28& 10:48:17.03& 12:38:27.3& 5019.6& 741& 11& 1.38& 25.91& 0.16&  44.8 (43)&  9.9&  4& 25.67\\
W29& 10:48:17.52& 12:37:10.5& 5021.4& 847& 11& 1.20& 26.06& 0.16&  45.7 (44)&  9.6& 57& 26.44\\
W30& 10:48:17.58& 12:38:07.8& 5018.8& 696& 15& 1.17& 26.09& 0.23&  23.3 (42)&  7.0&  5& 25.71\\
W31& 10:48:17.91& 12:38:09.3& 5018.6& 679& 20& 0.90& 26.37& 0.26&  17.5 (46)&  6.3& ..&  ... \\
W32& 10:48:17.98& 12:38:22.2& 5017.2& 599& 14& 1.04& 26.22& 0.21&  28.8 (45)&  7.9& ..&  ... \\
W33& 10:48:18.07& 12:38:44.0& 5018.4& 671& 11& 1.76& 25.65& 0.19&  31.4 (25)&  7.3& ..&  ... \\
W34& 10:48:18.12& 12:36:36.5& 5019.4& 729& 14& 1.36& 25.93& 0.21&  27.2 (25)&  7.2& 22& 26.06\\
W35& 10:48:18.58& 12:38:18.8& 5018.5& 676& 10& 1.51& 25.81& 0.17&  42.0 (43)&  9.1& 16& 25.98\\
W36& 10:48:18.83& 12:37:15.4& 5019.3& 721& 11& 1.09& 26.17& 0.21&  28.2 (41)&  8.2& ..&  ... \\
W37& 10:48:19.18& 12:37:30.5& 5018.4& 672& 18& 0.85& 26.43& 0.25&  19.8 (41)&  6.9& 50& 26.38\\
W38& 10:48:19.40& 12:38:16.4& 5017.3& 606& 16& 1.03& 26.23& 0.24&  22.4 (44)&  6.4& ..&  ... \\
W39& 10:48:19.64& 12:37:26.0& 5019.7& 747& 16& 1.10& 26.15& 0.23&  26.0 (42)&  6.6& ..&  ... \\
W40& 10:48:19.73& 12:36:44.5& 5020.7& 809& 11& 1.33& 25.95& 0.18&  38.6 (27)&  7.9& 44& 26.35\\
W41& 10:48:19.73& 12:38:16.1& 5016.9& 583& 10& 1.26& 26.01& 0.18&  38.5 (45)&  8.8& 40& 26.31\\
W42& 10:48:20.41& 12:38:03.5& 5018.2& 659& 16& 0.91& 26.36& 0.26&  18.8 (44)&  6.1& ..&  ... \\
W43& 10:48:20.50& 12:38:53.7& 5018.0& 643& 13& 1.52& 25.81& 0.23&  21.8 (19)&  6.5& 20& 26.04\\
W44& 10:48:20.75& 12:38:21.7& 5016.8& 572& 11& 1.29& 25.99& 0.18&  36.0 (45)&  8.6& 19& 26.01\\
W45& 10:48:20.88& 12:37:29.0& 5020.0& 766&  8& 1.42& 25.88& 0.14&  62.3 (45)& 10.8& 83& 26.73\\
W46& 10:48:20.92& 12:38:05.7& 5018.0& 648& 19& 1.00& 26.26& 0.22&  24.2 (44)&  6.6& 53& 26.41\\
W47& 10:48:20.93& 12:38:08.1& 5017.4& 607& 12& 1.03& 26.23& 0.23&  23.3 (43)&  6.8& ..&  ... \\
\enddata			       

\tablecomments{Col. (1) E and W denote PNe found in the East and West
fields, respectively; col. (2)--(3) right ascension (hms) and
declination ($\degr\arcmin\arcsec$), where the uncertainty in each
coordinate is $0\farcs4$ (includes random and systematic uncertainty);
col. (4) peak wavelength (\AA); col. (5) heliocentric line-of-sight
velocity ($\mr{km\ s^{-1}}$); col. (6) uncertainty in $v_\mr{helio}$;
col. (7) emission line flux ($10^{-16}\mr{\ erg\ cm^{-2}\ s^{-1}}$);
col. (8) emission line magnitude $m_{5007}\equiv -2.5\log(F)+13.74$
(no extinction correction was made); col. (9) uncertainty in
$m_{5007}$; col. (10) test statistic $S_\mr{LR}$ (number of data
points in spectrum); col. (11) signal-to-noise ratio; col. (12)--(13)
corresponding ID and $m_{5007}$ from \citet{cjf89}. }

\tablenotetext{a}{The spectra have a wide range in their number of
  data points and we determine $S_c(\alpha)$ for regions with one,
  two, or three pointing separately.  Hence, the values of $S$ listed
  in table~\ref{t:3384-pne} are not always directly comparable.}

\end{deluxetable}



\begin{figure}

\figurenum{\ref{f:n1549spectra}.2}
\plotone{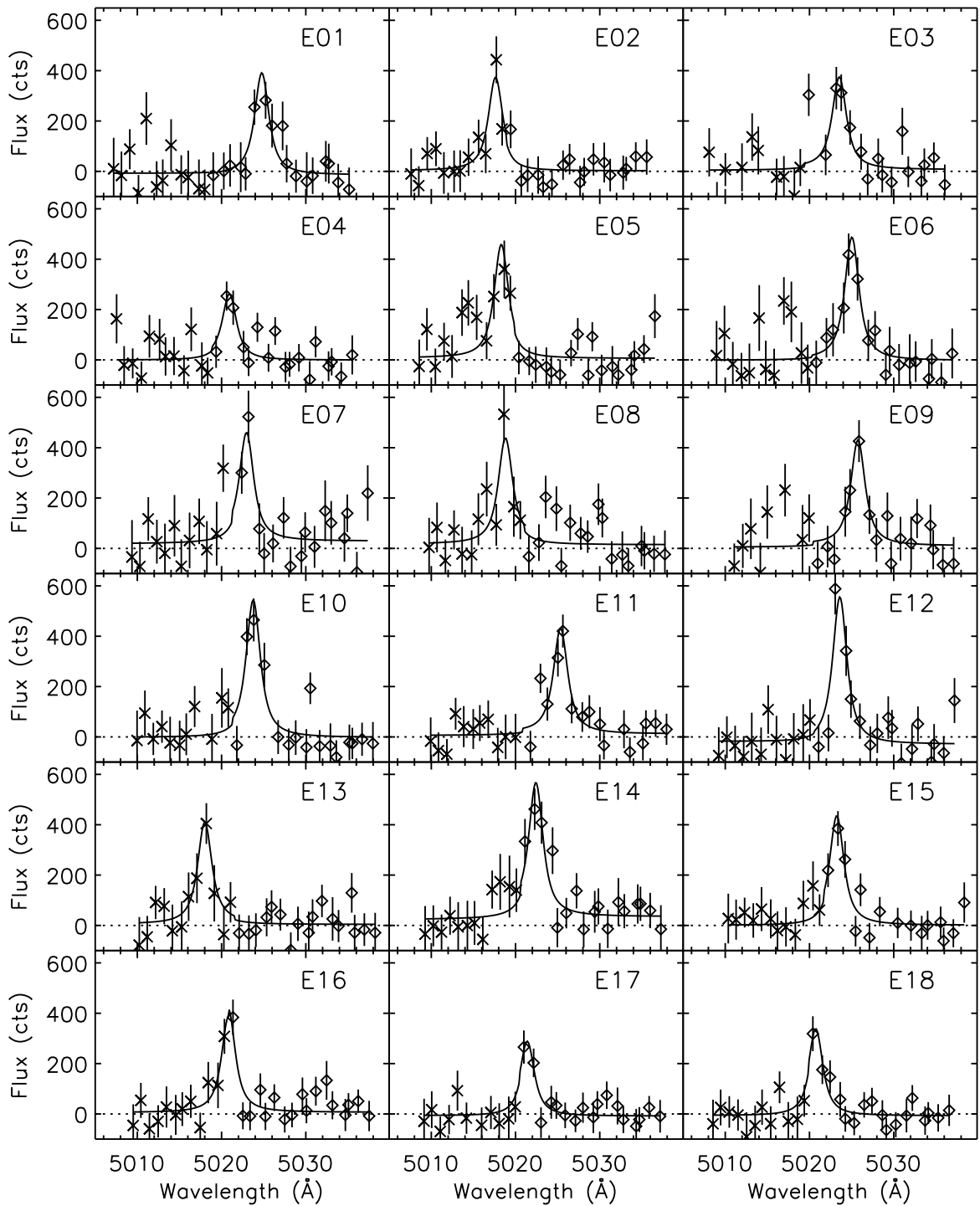}

\caption{RFP spectra of the NGC 3379 PNe.  Crosses denote observations
under the 5007-44 filter, squares the 5037-44 filter.  The curve is
the best fit line profile as described by equation~\ref{e:emmodel}.
The PN ID from Table~\ref{t:3379-pne} is shown in the top right corner
of each spectrum. \emph{Online only.}}

\end{figure}

\begin{figure}
\figurenum{\ref{f:n1549spectra}.3}
\plotone{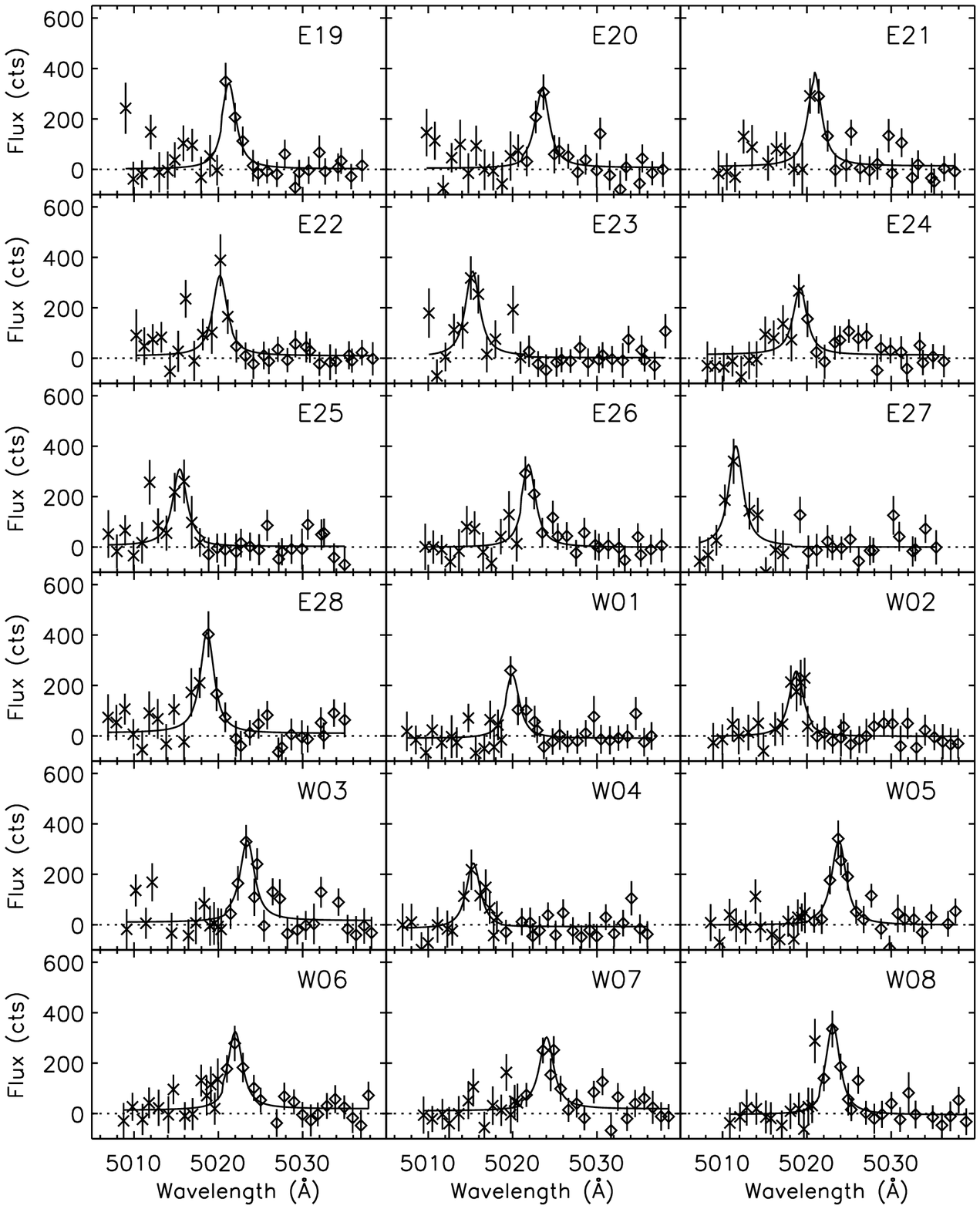}
\caption{Continued from Fig.~\ref{f:n1549spectra}.2.  \emph{Online only.}}
\end{figure}

\begin{figure}
\figurenum{\ref{f:n1549spectra}.4}
\plotone{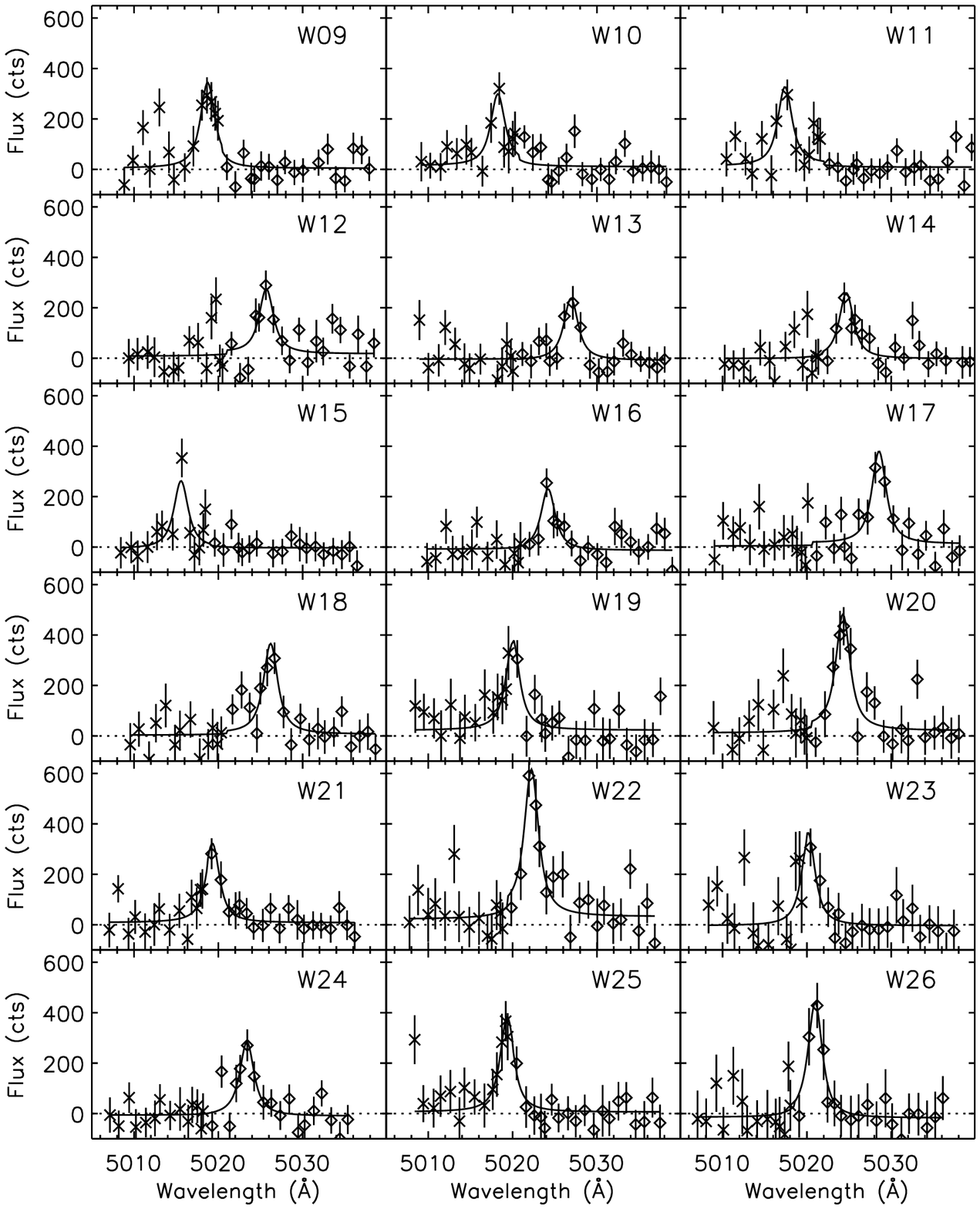}
\caption{Continued from Fig.~\ref{f:n1549spectra}.3.  \emph{Online only.}}
\end{figure}

\begin{figure}

\figurenum{\ref{f:n1549spectra}.5}
\plotone{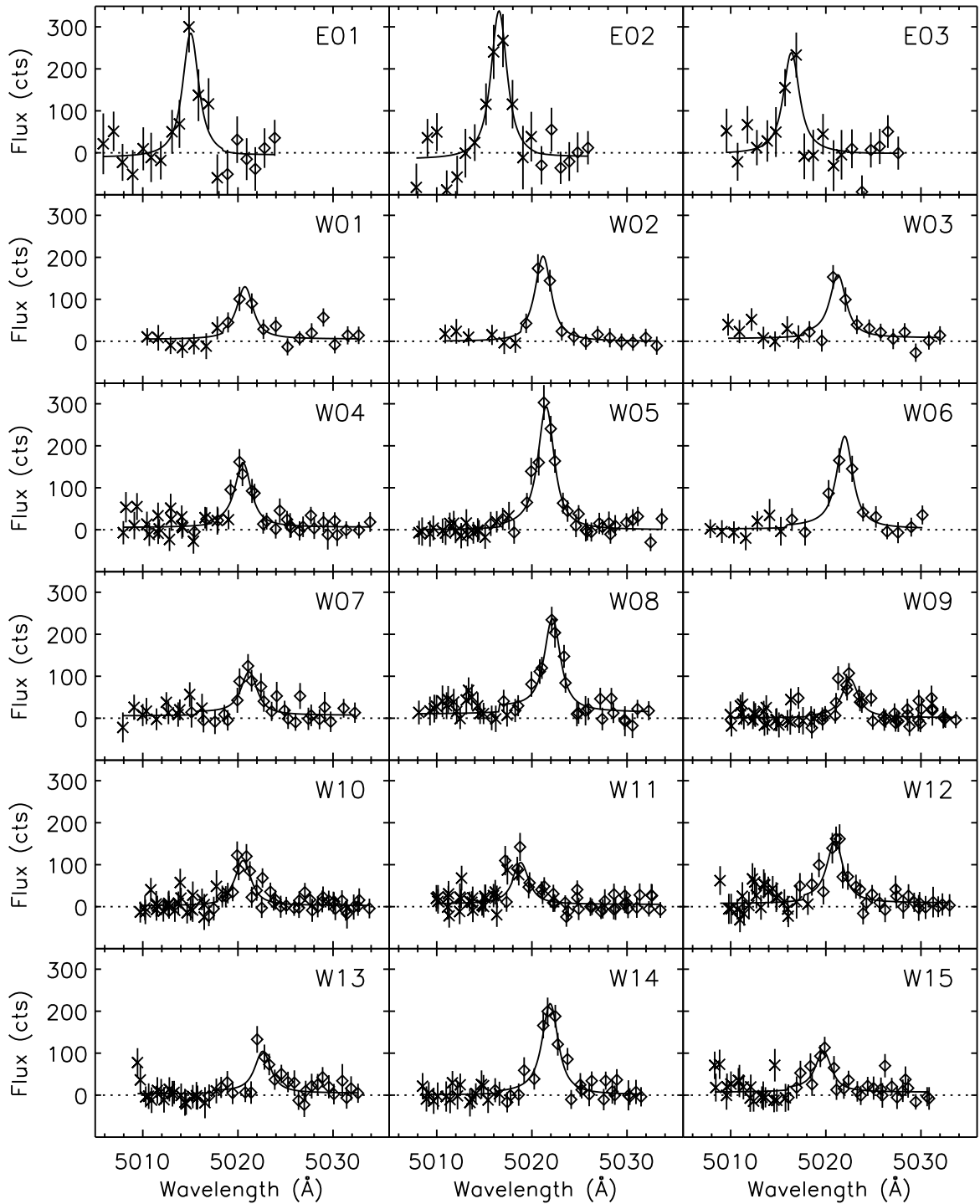}

\caption{RFP spectra of the NGC 3384 PNe.  Crosses denote observations
under the 5007-44 filter, squares the 5037-44 filter.  The curve is
the best fit line profile as described by equation~~\ref{e:emmodel}.
The PN ID from Table~\ref{t:3384-pne} is shown in the top right corner
of each spectrum. The PNe marked with E (W) were taken on the 1995
(1994) run and the flux scales are not directly comparible between the
two runs.  Note the changing ranges for the flux scales.  \emph{Online
only.} }

\end{figure}

\begin{figure}
\figurenum{\ref{f:n1549spectra}.6}
\plotone{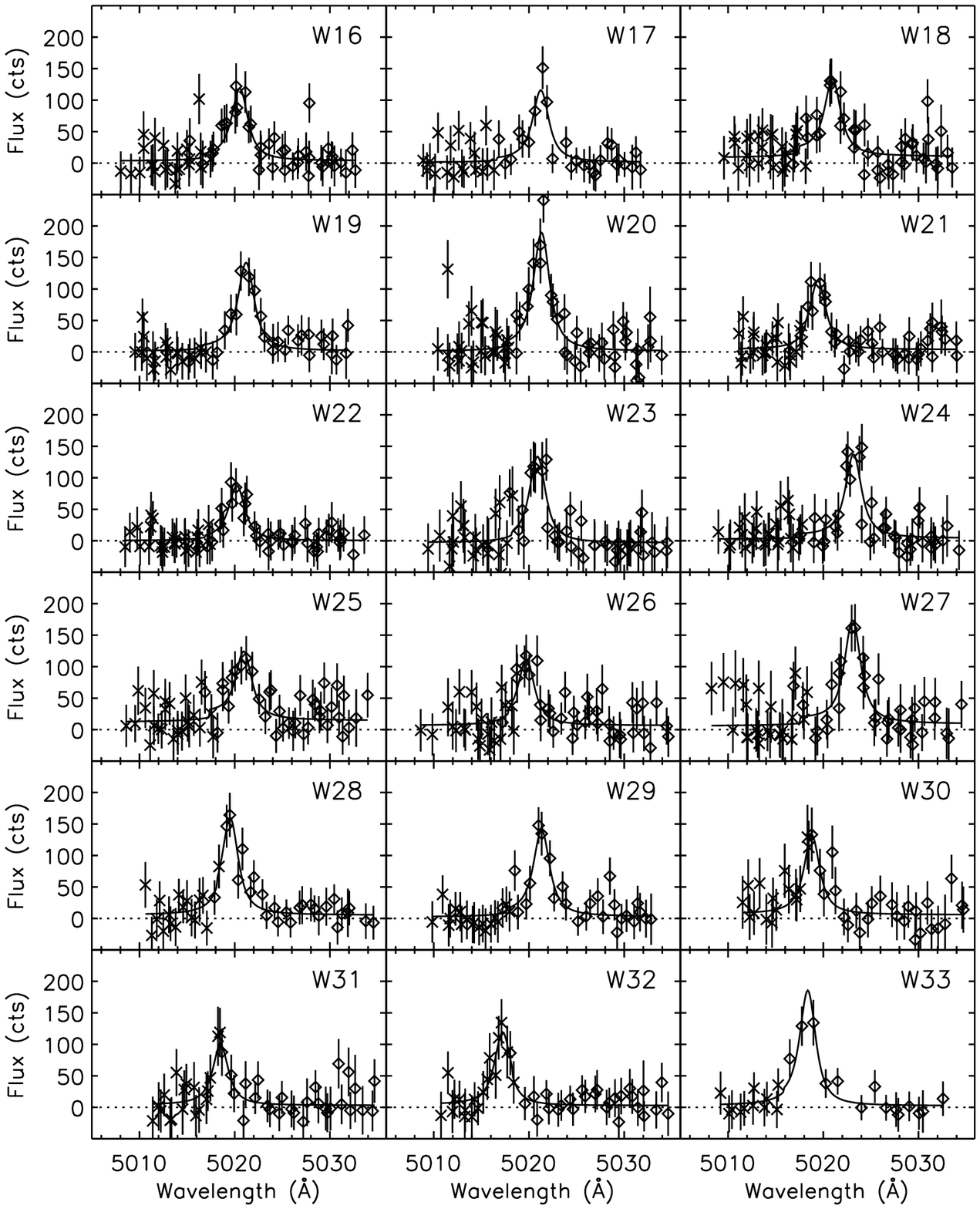}
\caption{Continued from Fig.~\ref{f:n1549spectra}.5.  \emph{Online only.}}
\end{figure}

\begin{figure}
\figurenum{\ref{f:n1549spectra}.7}
\plotone{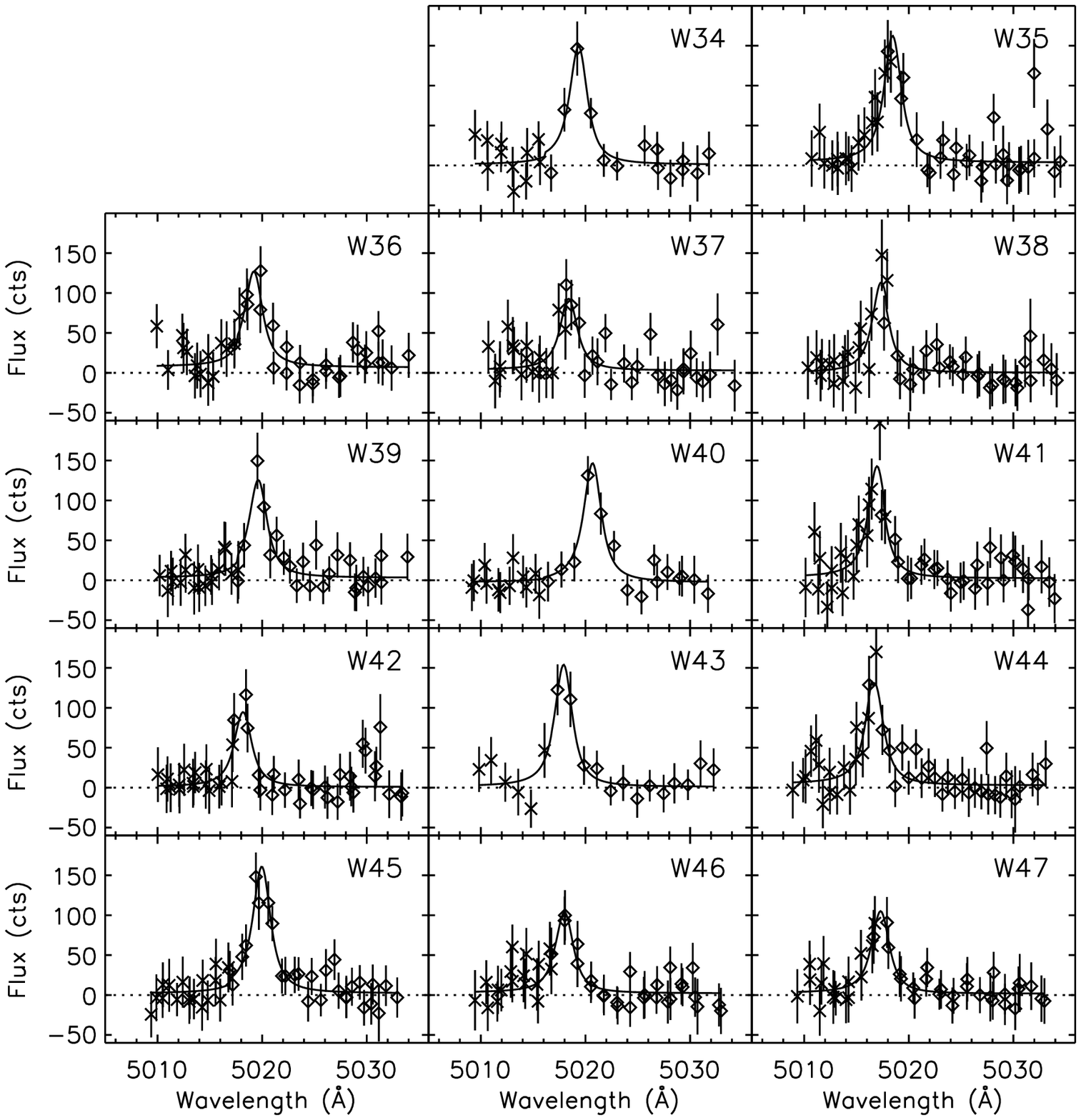}
\caption{Continued from Fig.~\ref{f:n1549spectra}.6.  \emph{Online only.}}
\end{figure}



\begin{figure}
\epsscale{0.45}
\plotone{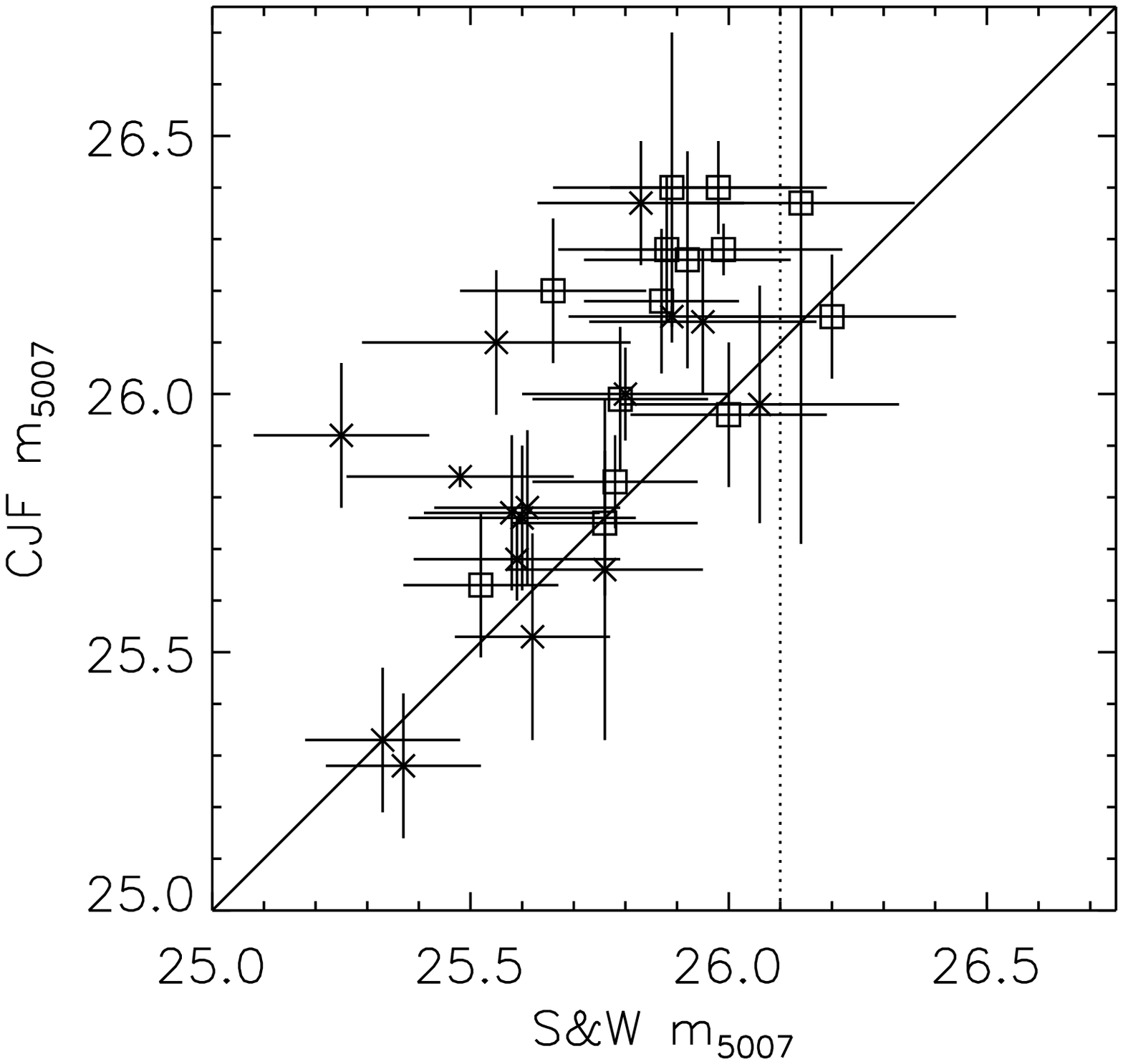}

\caption{Comparison of the $m_{5007}$ magnitudes of NGC~3379 in our
sample (S\&W) and those found by \citet[CJF]{cjf89}.  Unspecified
magnitude uncertainties in the latter sample were assumed to be
0.14~mag.  PNe in the East and West field are indicated by squares and
crosses, respectively.  The solid line indicates exact correspondence
between the two magnitude measurements; the dotted line the limiting
magnitude of our survey.}

\label{f:n3379magcomp}
\end{figure}



\begin{figure}

\plotone{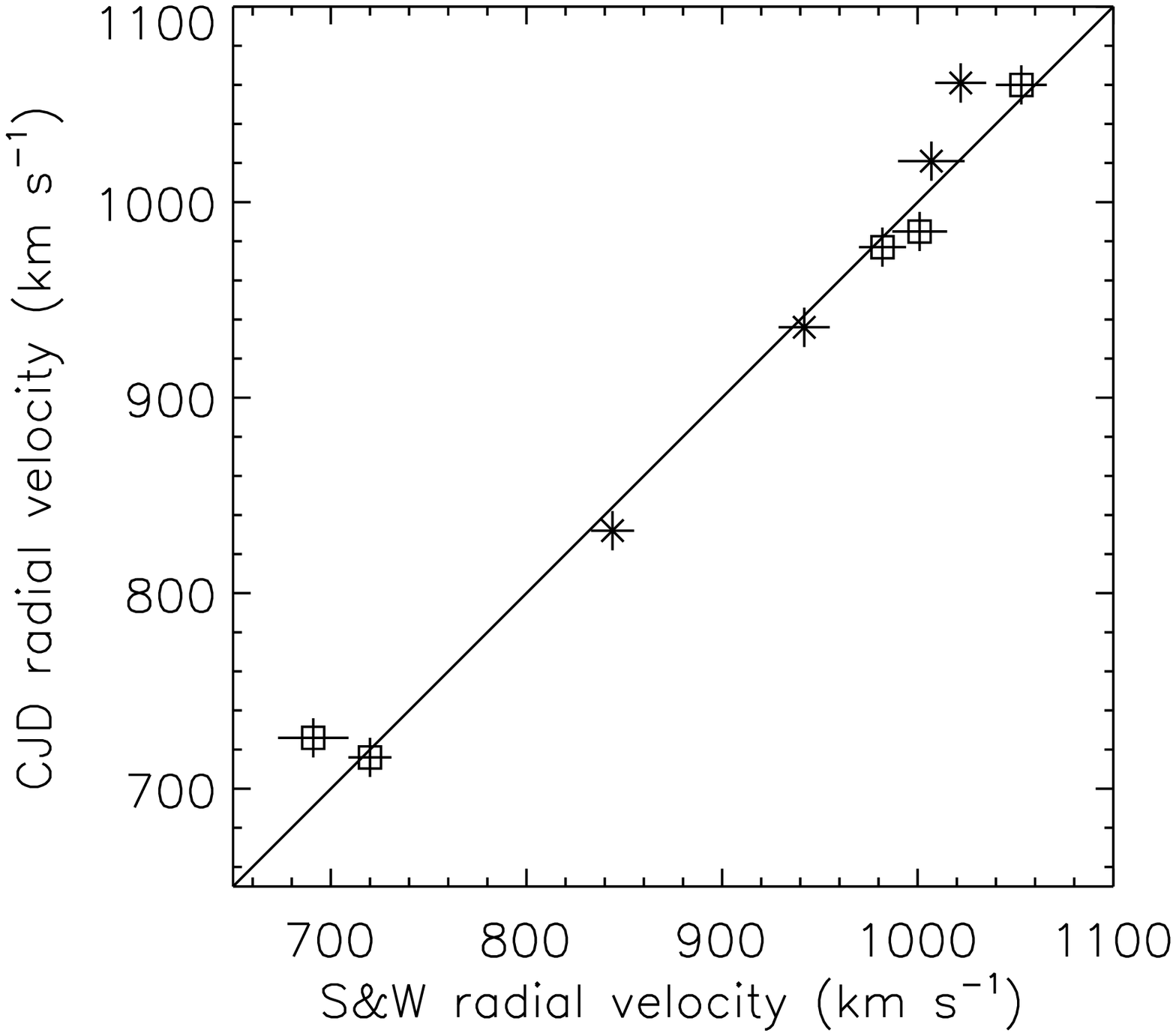}
\epsscale{0.45}
\caption{Comparison of the line-of-sight velocities of PNe in NGC~3379
determined by \citet[CJD]{cjd93} and those of our sample (S\&W).  We
assumed a a velocity uncertainty of $10\mr{\ km\ s^{-1}}$ for the
former.  PNe in the East and West field are indicated by squares and
crosses, respectively.  The solid line indicates exact correspondence
between the two velocity measurements. }

\label{f:n3379velcomp}
\end{figure}



\begin{figure}

\plotone{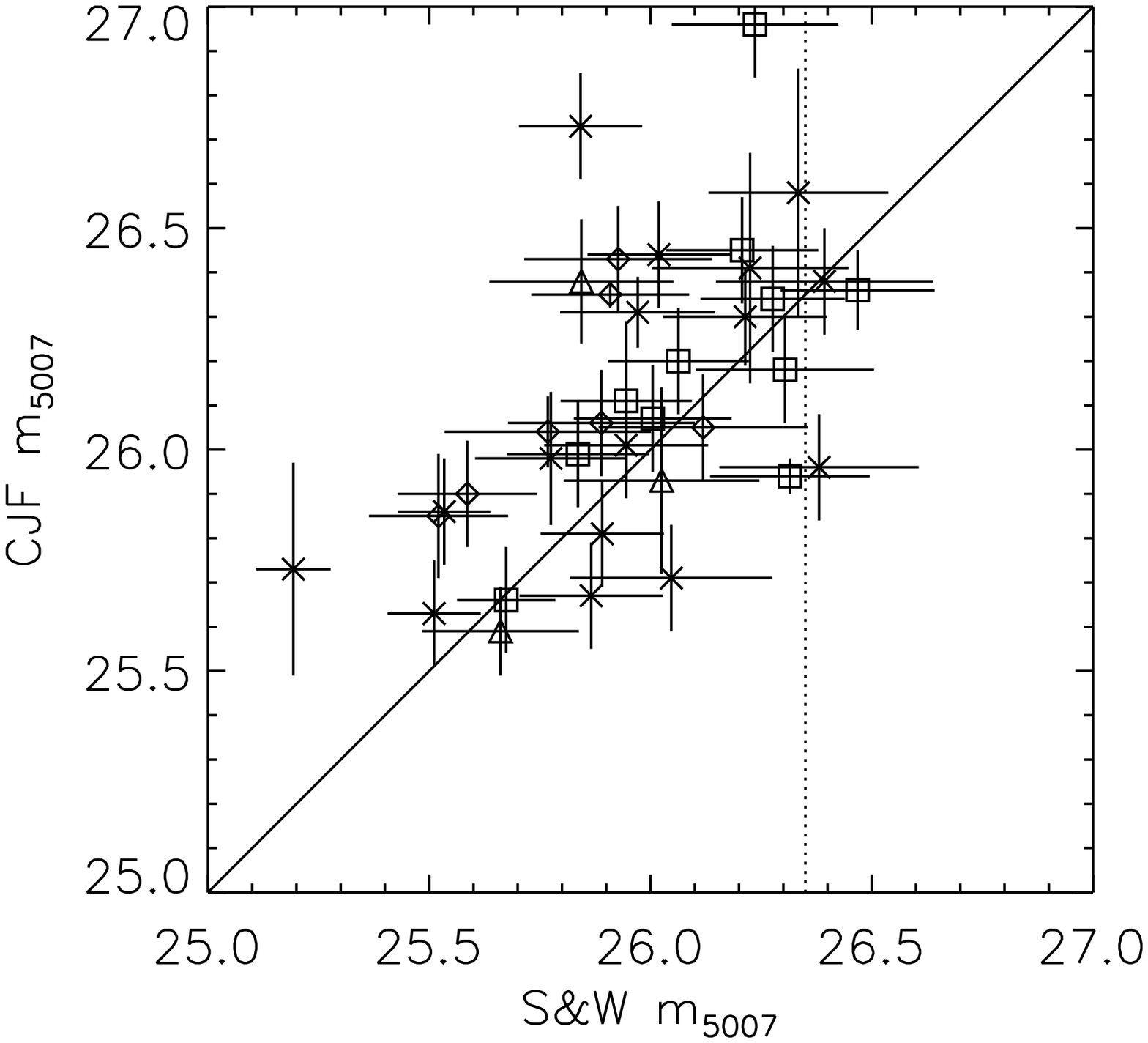}
\epsscale{0.45}
\caption{Comparison of the $m_{5007}$ magnitudes of NGC~3384 in our
sample (S\&W) and those found by \citet[CJF]{cjf89}.  Unspecified
magnitude uncertainties in the latter sample were assumed to be
0.12~mag.  The different symbols indicate the run and amount of data
points in the spectra: triangles - 1995 run (East field); diamonds,
crosses and squares - data from one, two or three pointings,
respectively.  The solid line indicates exact correspondence between
the two magnitudes.  The vertical dashed line is the average limiting
magnitude of our sample, but note that the limiting magnitude for the
1995 run is 25.9.}

\label{f:n3384magcomp}
\end{figure}



\begin{figure}

\epsscale{0.85}  
\plotone{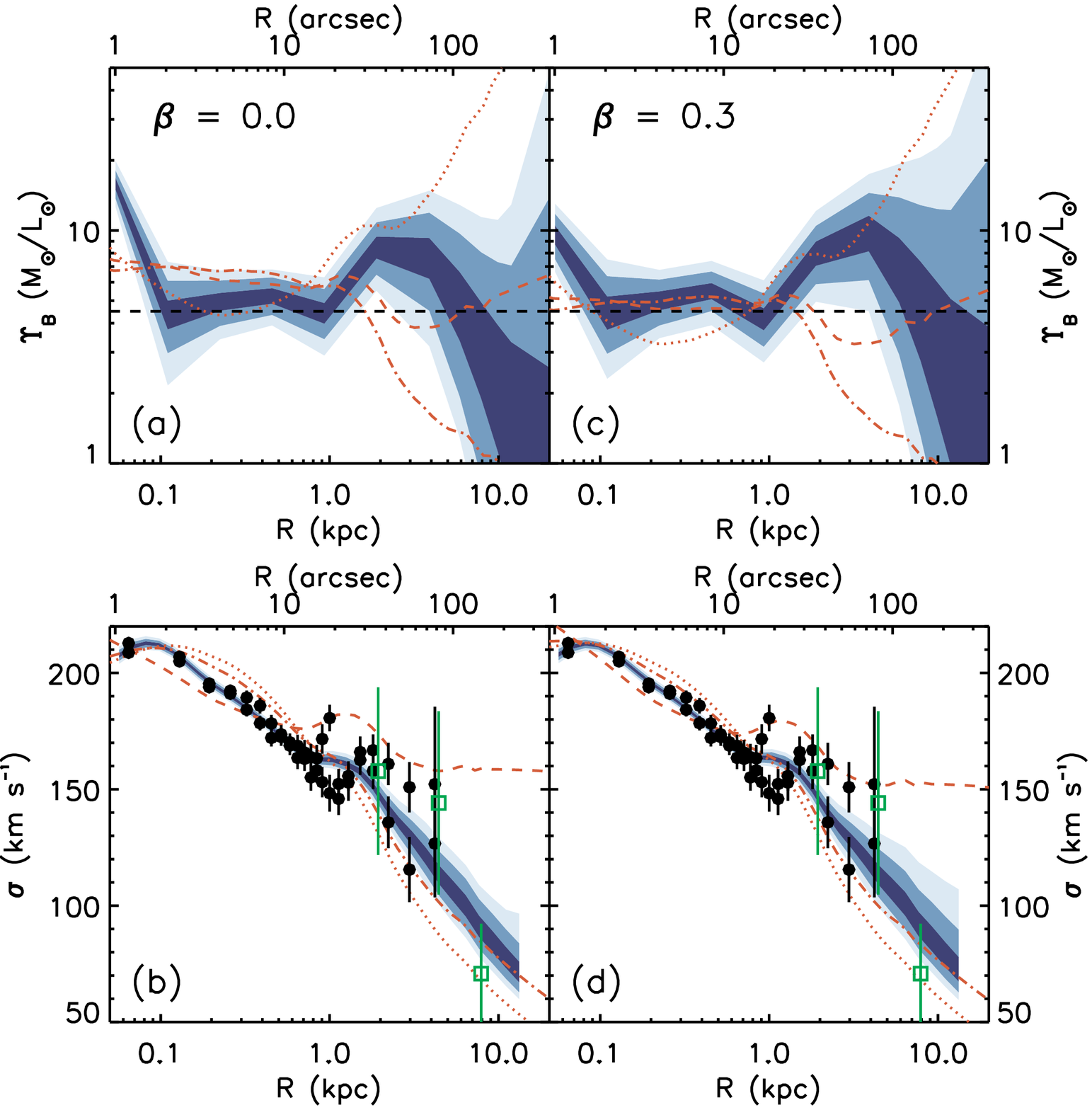}

\caption{Results from two MCMC runs for NGC~3379, one assuming
  isotropy ($\beta=0$, left column) and one assuming radial anisotropy
  ($\beta=0.3$, right column).  Credible regions (shades of blue) in
  each figure encompass, from the outside inwards, 99\%, 90\% and 50\%
  of the posterior distribution.  The red curves are the best fit
  results for a Hernquist model (dot-dashed), an NFW profile (dashed),
  and a pseudo-isothermal sphere (dotted).  Figures (a) and (c): the
  local mass-to-light ratio $\Upsilon(r)$.  The dashed line is the
  dynamical mass-to-light estimate from \citet{gerea01}.  Figures (b)
  and (c): the agreement with the data, where the filled symbols are
  the data from \citet{ss99} (error bars of less than $2.5\ \mr{km\
  s^{-1}}$ have been surpressed) and the green squares are the
  root-mean-square velocities of the PNe (this paper and
  \citet{cjd93}) in three bins with an equal number of PNe in each
  bin.  The binning is only applied for clarity in the figure and
  plays no role in our actual dynamical analysis.  }

\label{f:n3379_mcmc}
\end{figure}


\end{document}